MAXIMUM ENTROPY: THE UNIVERSAL METHOD FOR INFERENCE

by

Adom Giffin

A Dissertation

Submitted to the University at Albany, State University of New York

in Partial Fulfillment of

the Requirements for the Degree of

Doctor of Philosophy

College of Arts & Sciences

Department of Physics

2008

# Abstract

In this thesis we start by providing some detail regarding how we arrived at our present understanding of probabilities and how we manipulate them – the product and addition rules by Cox. We also discuss the modern view of entropy and how it relates to known entropies such as the thermodynamic entropy and the information entropy. Next, we show that Skilling's method of induction leads us to a unique general theory of inductive inference, the ME method and precisely how it is that other entropies such as those of Renyi or Tsallis are ruled out for problems of inference. We then explore the compatibility of Bayes and ME updating. After pointing out the distinction between Bayes' theorem and the Bayes' updating rule, we show that Bayes' rule is a special case of ME updating by translating information in the form of data into constraints that can be processed using ME. This implies that ME is capable of reproducing every aspect of orthodox Bayesian inference and proves the complete compatibility of Bayesian and entropy methods. We illustrated this by showing that ME can be used to derive two results traditionally in the domain of Bayesian statistics, Laplace's Succession rule and Jeffrey's conditioning rule.

The realization that the ME method incorporates Bayes' rule as a special case allows us to go beyond Bayes' rule and to process both data and expected value constraints simultaneously. We discuss the general problem of non-commuting constraints, when they should be processed sequentially and when simultaneously. The generic "canonical" form of the posterior distribution for the problem of simultaneous updating with data and moments is obtained. This a major achievement since it shows that ME is not only capable of processing information in the form of constraints, like MaxEnt and information in the form of data, as in Bayes' Theorem, but also can process both forms simultaneously, which Bayes and MaxEnt cannot do alone.

We illustrate some potential applications for this new method by applying ME to potential problems of interest.

# Acknowledgments

This thesis is dedicated to Mrs. Linda Flynn who is the most altruistic person I know. She is the only one that knows how far I have come in order to write this today. I will be forever indebted to her. I would also like to dedicate this to Mr. J.R. Elmleaf, who treated me like an equal when outside of class.

There are a lot of people that I need to thank that helped me get to Albany. First, of course, I need to thank my beautiful wife Traci for supporting me and putting up with my insane ravings for the last five years. I thank my mother for helping with both the transition of moving out here and now finishing up here. I need to thank all the professors that wrote letters for me, especially Professor Maria Vaz whose letter helped me get a job when I really needed one. Thank you to Jill Kurcz for her great kindness in helping me paint my house so I could sell it and for general encouragement.

There are many professors at Albany that I must thank. First and foremost I thank my advisor, Professor Ariel Caticha, without whom I would have no thesis. Some of the more fantastic discussions he indulged me in became collaborations that have open many doors for me along the way. I cannot express enough thanks to Professor Kevin Knuth for his selfless assistance, support and most of all friendship. I would like to thank Professor John Kimball for the opportunity to teach among other things such as help with comprehensive exam questions and serving on my thesis committee. I would like to thank Professor Keith Earle for helpful suggestions regarding my thesis, for being on my both my oral and thesis committees and general assistance. Thanks to Professor Carlos Rodríguez for discussions and being on my thesis committee as well.

Finally, I must thank the friends and colleagues that I have made along the way here, especially Dr. Carlo Cafaro who, besides being a good friend, helped me academically many times and Professor Arnold Zellner, whose fame and notoriety does not diminish his genuine graciousness. Last but not least I must thank the department administrator, Ms. Donauta Collins, who helped me out with a lot of little things that would have been very big things for me.

# Contents

# Preface

It is well established and agreed upon why we use probabilities; we use probabilities because we lack complete information. However, 'how' they are used has been a contentious problem. How we use probabilities depends on one's definition of the word.

Every practically useful question that we can ask is one in which we lack information. Questions we may ask are, "Is there life on Mars?", "Is the price of gold going to drop tomorrow?" and "What is the cure for cancer?" Of course the questions can be much more mundane, such as, "How long is my pen?" or "Should I cross the street?" The feature that all of these questions share is that in each we lack information to some degree. We can only do one thing when faced with a lack of information: reason by *induction*.

In all cases, the object is to process information that we have regarding a certain question in a systematic and consistent way, in order to provide the best answer to the question. The Maximum relative Entropy (ME) method that we detail in part two of this thesis turns out to be the *universal* method for processing information. We will prove this using general examples with information in the forms of constraints and data. However, it important to point out in the beginning that while the examples may be generic the method is *universally applicable*.

# Part 1: A brief history of probability and entropy

We will attempt to give an extremely brief overview of the main lines of logic and highlight what we feel are the key ideas that have brought us to our current understanding of probabilities and how to process information. We will supply references to which the reader may further research the foundations that have lead to the work presented here.

# Chapter 1

# Probability

The concept of probability is as old as time itself. For humans, it is almost impossible to communicate without using the words, 'probably', 'likely', etc. The reason is obvious; we simply lack enough information to answer most questions. But why ask questions in the first place? Because we wish to learn. Question and answers are the foundations of learning. Therefore, probabilistic notions are *necessary* to learn. The process of learning *is* the process of inference. We learn by processing partial information which yields answers with uncertainty – probabilities. Additionally, when we get new information, we add it to what we already know, reprocess and arrive at new probabilities.

However, probabilities by themselves are practically useless. It is when we *compare* probabilities that they become *useful*. For example: suppose that there are three paths and one must make a decision on which path to follow. Suppose that one gains sufficient information regarding a particular path where the probability of survival along the path is 30%. Should one take that path? It depends upon the probabilities of other paths. If one of the alternatives has a higher probability of survival then the answer is no. One may assign equal probabilities to the other paths if no other information is present, but this is still so that we can compare them. In this chapter we will attempt to briefly describe how we have come to our present understanding of probabilities.

## 1.1 James (Jacob) Bernoulli (1654–1705)

Although probabilities had been examined prior to Bernoulli, he was the first one to use a common language to describe them – mathematics. Bernoulli gave his greatest work (which was published 8 years after his death) the name, *Ars Conjectandi* (The Art of Conjecturing) [1]. Literally the art of guessing. However, his guesses were meant to be *consistent* and more importantly his *process* of guessing was meant to be consistent. In other words, there are both wise ways of guessing and foolish ways of guessing. This is the single most important key to learning. It will be the overriding theme throughout this thesis: if a process of guessing or *inference* is inconsistent, then it is foolish.

One of the first ideas that he proposed was that if one did not know anything about various possible outcomes, then one should assign each outcome the same probability. This was called the "principle of insufficient reason" (PIR).[1] For example, if two dice that we know nothing about are thrown, the probability of getting any combination of numbers is equally likely. In this case, there are 6 ways out of 36 possibilities to get a 'seven'.[2] To make this more general, we look at the hypothesis space, $H$, which consists of our potential propositions, $A$, so that $H = H(A_1 \ldots A_k)$. This is sometimes called the field of discourse. We will label the number of cases where $A$ is favorable, $M$, and label the total number of cases, $N$, so that

$$p(A) \equiv \frac{M}{N}, \tag{1.1}$$

where $p(A)$ is the probability of getting $A$, or of $A$ being true. For our example, the probability of getting a 'seven' with two dice is $1/6$.

---

[1] Contrary to a great number of authors, PIR was coined by Johannes von Kries [4].
[2] PIR was in its infancy at this stage and only used by Bernoulli in cases of symmetry, like dice.

This result seems very reasonable. However, one may not see the problem with (1.1) immediately. The problem is that one would have to know *all* of the cases, $N$ to assign a probability. He realized that (1.1) was not the solution for practical matters, only for games of chance. To quote the man himself, "...what mortal will ever determine, for example, the number of diseases...? It would clearly be mad to want to learn anything this way."

For a more pragmatic solution, it seems reasonable that we could extend (1.1) by interpreting the number of favorable *observed outcomes* (trials) as favorable cases and total outcomes as total cases thus

$$\nu(A) \equiv \frac{m}{n}, \tag{1.2}$$

which is the frequency of observed favorable occurrences where $m$ is the number of observed favorable outcomes and $n$ is the total observed outcomes. Yet there is another problem. The new problem is that this is simply a ratio of observed cases and not necessarily what one should *expect* to observe as outcomes from a different number of trials. In other words, in 36 rolls should we expect to get 6 'sevens'? If 1000 people rolled perfectly symmetric dice, would they all get exactly 1/6 of their roles to come up 'seven'? To put this in more technical terms, if we considered (1.2) a 'probability' should we expect observe it in *any* population? It was well understood by his time that as one observed more outcomes, one decreased the "danger there will be of error" [2]. Therefore, Bernoulli sought to determine a relationship between (1.1) and (1.2) which he showed by penning his famous formula,

$$P(m|p,n) = \binom{n}{m} p^m (1-p)^{n-m} , \tag{1.3}$$

where $P(m|p,n)$ is the probability of getting $m$ favorable *observed* outcomes, *given* the ratio $p$, and the number of total outcomes, $n$. This is known as the Binomial Distribution. Where $1-p$ represents all non-favorable outcomes. In our example, all number combinations that are not 'seven'.

We can see the relationship between $p$ (1.1) and $\nu$ (1.2) directly by the following: Consider the expected value of the number of success, $\langle m \rangle$ where our probability function is $P(m|p,q,n) = \binom{n}{m} mp^m q^{n-m}$,

$$\langle m \rangle_{p,q} = \sum_{m=0}^{n} mP(m|p,q,n) = \sum_{m=0}^{n} \binom{n}{m} mp^m q^{n-m} . \tag{1.4}$$

We substitute

$$mp^m = p \frac{\partial p^m}{\partial p}, \tag{1.5}$$

so that

$$\langle m \rangle_{p,q} = p \frac{\partial}{\partial p} \sum_{m=0}^{n} \binom{n}{m} p^m q^{n-m} . \tag{1.6}$$

Next we note the binomial *theorem* without proof,

$$\sum_{m=0}^{n} \binom{n}{m} p^m q^{n-m} = (p+q)^n \ . \tag{1.7}$$

Substituting this into our expectation (1.4),

$$\langle m \rangle_{p,q} = p \frac{\partial}{\partial p} (p+q)^n = pn(p+q)^{n-1} \ . \tag{1.8}$$

Specifically we wish to know $\langle m \rangle$ when $q = 1 - p$. With this substitution, our final result is,

$$\langle m \rangle_{p,1-p} = pn \quad \text{or} \quad \frac{\langle m \rangle}{n} = \langle \nu \rangle = p \ , \tag{1.9}$$

which shows that it is the *expected value* of the frequency that is equivalent to the probability. Thus, in general, the probability is *not* equivalent to the frequency. However, by a similar set of calculations, one can show that the expected value of the variance is

$$\langle (\nu - p)^2 \rangle = \frac{p(1-p)}{n} \ . \tag{1.10}$$

As $n$ tends to infinity, the variance tends to zero. Therefore, as Bernoulli proposed, the frequency equals the probability in the limit,

$$\lim_{n \to \infty} (\nu) = p \ , \tag{1.11}$$

which he called his "Golden Theorem". In 1835, his student, Siméon-Denis Poisson (1781–1840), rewrote his theorem and called it "La loi des grands nombres" (The law of large numbers) [3]. It should be noted that many people after him worked on this including Chebyshev, Markov, Borel, Cantelli and Kolmogorov, all of whom helped to transform the law into two forms: the weak law and the strong law.

Having shown that probability is not equivalent to frequency, except as the number of trials tends to infinity, we get to Bernoulli's big problem: "How does one determine the probability without being able to observe an infinite amount of trials?" His definition of probability (1.1) was not practical and the frequency (1.2) could never be the *true* probability. Therefore, he concluded that for practical problems, one could only use the frequency as an estimate of the true probability. This 'estimate' would used for another 150 years by the majority of mathematicians and statisticians. Yet it was still to be resolved what to do when one had a few samples, if any at all.

## 1.2 Thomas Bayes (1702–1761)

In one sense, the Reverend Thomas Bayes' contribution would have been too small to have his own section in this work. On the other hand, it would be strange not to have a section dedicated to the man whose name is given to one of the most fundamental and influential theorems of all time.

Besides being a British clergyman, Bayes was also an amateur mathematician. Amongst his writings was something he called, "inverse probability" [5]. Essentially this was an attempt to solve the problem that Bernoulli did not; how to determine the probability of something *given* its frequency. In other words, given $m$ and $n$, what was $p$? Or to state it formally using Bernoulli's notation, he found that the probability that (1.1) lies in the interval $p < p(A) < p + dp$ is

$$P(dp|m,n) = \frac{(n+1)!}{m!\,(n-m)!} p^m \left(1-p\right)^{n-m} dp \; . \tag{1.12}$$

Today this is called the Beta distribution. Unfortunately, Bayes notes were obscure and muddled. As a result, it is not clear what his motivations or intentions were with the above solution. Aside from (1.12), Bayes did little more to contribute to the statistical methodology that bears his name.

## 1.3 Pierre-Simon Laplace (1749–1827)

"It is remarkable that a science which began with the consideration of games of chance should have become the most important object of human knowledge." – Laplace

In almost his first published work, Laplace rediscovered Bayes' solution to the 'inverse problem' and restated it with far greater clarity than Bayes. Laplace generalized the inverse problem in the following way: Suppose one observes a set of events, $X = \{x_1 \ldots x_n\}$. Let the set of possible causes of said events be, $\Theta = \{\theta_1 \ldots \theta_N\}$. Let us also assume that we know the probability of observing the set of events given a particular cause, $P(X|\theta_i)$.[3] Finally, we initially consider all of the causes 'equally likely'.[4] Laplace then proposed that the probability for getting a particular cause,

---
[3]This is called a *conditional* probability, as the probability of $X$ being true is conditional or dependent on $\theta_i$ being true. It also should be noted that this is a modern notation.

[4]Equally likely means that we consider the prior probabilities to be *uniform*.

given a set of events, could be written as,

$$P(\theta_i|X) = \frac{P(X|\theta_i)}{\sum_{j=1}^{N} P(X|\theta_j)} \ . \tag{1.13}$$

Later, Laplace then generalized this further by suggesting that the probability $P(X|\theta_i)$, could be weighted by other information, such as prior information about the cause. He would write this as $P(\theta_i|I)$, where $I$ is some initial information, such as a die having $n$ tosses. Adding this weight we get what is called Bayes' Theorem,

$$P(\theta_i|X,I) = \frac{P(\theta_i|I)P(X|I,\theta_i)}{\sum_{j=1}^{N} P(\theta_j|I)P(X|I,\theta_j)} = \frac{P(\theta_i|I)P(X|I,\theta_i)}{P(X|I)} \ , \tag{1.14}$$

where $P(X|I)$ is often referred to as the evidence.

Laplace was not without his own problems as well. The first and most obvious problem is what $P(\theta_i|I)$ *is* and *how* do we acquire it? Laplace suggested that if we are ignorant of the prior information regarding $\theta_i$ and therefore do not know the prior probability, $P(\theta_i|I)$, we should assume $P(\theta_i|I)$ is uniform, i.e., $P(\theta_i|I) = 1/N$. In other words, use the principle of insufficient reason, just as Bernoulli suggested. Unfortunately, this puts us right back where we started from; why should we assume this is true? Laplace restates the principle thusly, "The theory of chance consists in reducing all the events of the same kind to a certain number of cases equally possible, that is to say, to such as we may be equally undecided about in regard to their existence, and in determining the number of cases favorable to the event whose probability is sought" [6]. This produced even more criticism. When should we use PIR? When there is a symmetry, as in so called fair dice, as Bernoulli suggests? Or when we are 'undecided'? Meaning when we are ignorant? Laplace offers no explanation or suggestion. To make matters worse, he simply states (1.14) to be correct without rigorously showing the path of his logic [7]. Thus, despite his enormous success in fields such as astronomy, meteorology, geodesy, population statistics, etc., he was only pragmatically successful using (1.13) and not (1.14). Thus, Bayes Theorem went unused and heavily criticized.

We conclude with Laplace by discussing one of the most fundamentally brilliant suggestions that he makes in his writings. At this time (and long after) probabilities were thought of as a property of the object. For example: A particular roulette ball would have a certain probability of settling into the 17 black slot. This is true for every observer if the roulette wheel remains untouched. However, another ball on another roulette wheel would have its own probability of landing in 17 black, so the common logic went. Laplace suggested that it is not the fact that the ball is different but that the *knowledge* or information about the ball is different. Thus if the information regarding the balls were the same, then both balls would have the same probabilities. Furthermore, he was criticized because his use of "we" in his statement above regarding his theory of chance. It suggests that different people could have different probabilities for the same object. This is precisely what he means! Again, the probability is a product of the information *regarding* the object, not an artifact of the object itself. Thus two people who have different pieces or amounts of information could indeed assign different probabilities to the same object. This is another key point in this thesis. To enhance Laplace's idea we turn it around and state it thusly: Two rational agents, who use the *same process* to assign probabilities to the same event, and have the *same information* regarding the event, *must* assign the *same* *probabilities* to the same event. This is a result of *consistency*.

## 1.4 Richard Threlkeld Cox (1898–1991)

In 1946, Cox wrote a small yet monumental paper that finally gave Laplace's ideas flesh [8]. In it, he gave examples of common inference problems that sampling theory could not handle. Yet, he noted, the goal of both the frequentists and the Bayesians were "universal". In other words, both want to do the same thing: infer something given partial information. Also, both views expect the probabilities to follow a basic rule of logic: If the probability of $A$ is greater than the probability of $B$, and the probability of $B$ is greater than $C$, then the probability of $A$ should be greater than that of $C$. This is called *transitivity*. We wish to rank them so that we can compare them. Therefore, he asked the question, "Is there a *consistent* mathematically derivable way to assign probabilities?" The answer to this was indeed yes and Cox applied mathematical rigor on Laplace's logic. To do this Cox makes one assumption: a proposition is either true or false. From this he applies the mathematical structure which can be contained in two axioms:

**Axiom 1:** The plausibility that a proposition $a$ given $c$ is true must be related to the negative of the proposition, *not a* or $\tilde{a}$, given $c$,

$$\mathcal{P}(\tilde{a}|c) = f(\mathcal{P}(a|c)) , \qquad (1.15)$$

where $f(.)$ is a monotonic function, $\mathcal{P}(a|c)$ represents our belief in $a$ given $c$, $\mathcal{P}(\tilde{a}|c)$ represents our belief in $\tilde{a}$ given $c$ and there exists a function $f(.)$ that transforms

$\mathcal{P}(a|c)$ into $\mathcal{P}(\tilde{a}|c)$. This is also known as the negation relation.

**Axiom 2:** The plausibility that the proposition $a$ and proposition $b$, given $c$ are true must be related to plausibility that proposition $a$ given $c$ is true and the plausibility of proposition $b$ given $a$ and $c$ is true,

$$\mathcal{P}(a,b|c) = g(\mathcal{P}(a|c), \mathcal{P}(b|a,c)) , \qquad (1.16)$$

where $g(.)$ is the monotonic function of transformation.

With these two axioms, we use Cox's First Theorem of Regraduation to attain,

$$p(a,b|c) = p(a|c)p(b|a,c) , \qquad (1.17)$$

and his Second Theorem of Regraduation to attain,

$$p(a|c) + p(\tilde{a}|c) = 1 , \qquad (1.18)$$

where we now call $p(.)$ a probability in the Laplacian sense of the word. For brevity we write the result without proof. However, we point the reader to [8] where Cox clearly and thoroughly wrote his proof in detail. The result (1.17) is called the *product rule* and (1.18) is called the sum rule or *normalization rule*. We can arrive at a relation similar to the Kolmogorov sum rule [9] by using the product and normalization rule along with the relation,

$$\widetilde{(a+b)} = \tilde{a}\tilde{b} , \qquad (1.19)$$

which says that if the statement $(a$ or $b)$ is FALSE then the both $a$ and $b$ must also be FALSE. The proof is trivial,

$$\begin{aligned}
p(a+b|c) &= 1 - p(\tilde{a}\tilde{b}|c) = 1 - p(\tilde{a}|c)p(\tilde{b}|\tilde{a},c) = 1 - p(\tilde{a}|c)(1 - p(b|\tilde{a},c)) \quad (1.20)\\
&= p(a|c) + p(\tilde{a},b|c) = p(a|c) + p(b|c)p(\tilde{a}|bc) = p(a|c) + p(b|c)(1 - p(a|bc))\\
&= p(a|c) + p(b|c) - p(a,b|c).
\end{aligned}$$

which is often called the *sum rule* where (1.18) is simply called the normalization rule.

There are two main reasons why these derivations of the sum and product rules are so powerful. First, there is no mention of priors or frequencies. Second, they are a result of mathematical consistency. In other words, when a result can be calculated two different ways, the results *must* agree. Therefore, if one wishes to use probabilities for the purposes of inference, then the *only* rules that can be used to manipulate those probabilities are the sum and the product rule or regraduations thereof. Any other rules would be inconsistent. We can now write,

$$p(a|c)p(b|a,c) = p(b|c)p(a|b,c) , \quad (1.21)$$

due to the symmetry in the 2$^{\text{nd}}$ axiom. Dividing both sides by $p(a|c)$ yields,

$$p(b|a,c) = \frac{p(b|c)p(a|b,c)}{p(a|c)} , \quad (1.22)$$

which is Laplace's "Bayes' Theorem". Therefore, Bayes' theorem is a result of mathematical consistency. We apply modern labels to the pieces by calling $p(b|a,c)$ the *posterior*, $p(b|c)$ the *prior*, $p(a|b,c)$ the *likelihood* and $p(a|c)$ we call the *evidence* or the normalization.

It should be noted that all the results in this section on Cox could be written without the probabilities and functions conditional on $c$. For example Bayes' Theorem could be written as

$$p(b|a) = \frac{p(b)p(a|b)}{p(a)} \ . \tag{1.23}$$

However, the 'extra' conditionalization throughout is intended. Probabilities only have meaning when they are in a specific context. To borrow somewhat from Aristotle; probabilities abhor a vacuum.

## 1.5 Summary

We started this chapter suggesting that the reason we use probabilities is because we wish to learn. So far we have only considered how to assign probabilities consistently, but not about learning from new information. Laplace said that the 'weight' he assigned to the likelihood was due to prior information. Thus, he did not only generalize Bernoulli, he attempted to generalized *how we learn*. The weight or old information (the prior) is a probability that reflects old information. We take this old probability and multiply it by the likelihood (our model with data put into it) in order to get a new probability assignment, or an *updated* probability (after normalization).

However, there are two problems. The first is that we still have not justified the use of PIR to assign prior probabilities. The second is that we get information in forms other than data. For example, in thermodynamics, the key to determining the probability distribution for a gas in thermal equilibrium is knowing that the relevant information is the expected energy. Yet this is not data. Therefore, how do we update our probabilities with information in other forms? The solution to the first problem comes in the next chapter and the solution to the second, a further generalization of how we 'update' our probabilities, comes in chapter 4.

Finally, we must address the confusion regarding terminology such as, 'degrees of belief', subjective and objective probabilities. Some do not like 'degrees of belief' as describing probabilities because they think, "How can what you 'believe' affect the outcome concerning a physical object?" In a similar vein, "How can the probability of a physical object be subject to a particular person?" The confusion rests purely in the connotation of the words being used. Let us give our definition of a probability:

**Definition:** A *probability* is an *abstract* notion that represents the *degree* of *plausibility* of a proposition, *subject* to *information* regarding that proposition.

It is abstract because it is a representation of our uncertainty. The object is certainly going to do something; it's just that we will never be *certain* what that 'something' will be. It is a 'degree' because, as we have already mentioned, it is when we *compare* probabilities that they have meaning. Also, we want the degrees to be transitive so that we can compare them *consistently*.

Therefore, by using 'belief' instead of 'plausibility' or 'subject' instead of 'given', we do not change how probabilities are used. Notice, they are subject to the information, not the people. Of course, it is very reasonable that people who receive different information may assign different probabilities. However, if they have the *same* information (same data, model and prior) and use the *same* method to process that information then they *must* calculate the same probabilities. Any other way is inconsistent and thus, useless.

We note that others continue to explore alternative approaches, such as lattice theory, to explain probabilities [10].

# Chapter 2

# Thermodynamic Entropy vs Statistical Mechanics

Boltzmann extends the subject thermodynamics, by introducing probabilities. Gibbs called this extended subject "statistical mechanics" and cleaned up some issues with Boltzmann's derivations. However, each of the participants were reading each other's papers, which makes it difficult to determine who did what and when. Boltzmann often (unfairly) gets the lion share of credit in most textbooks. Here we again highlight the key concepts of how these ideas were introduced and came about.

## 2.1 Ludwig Edward Boltzmann (1844–1906)

Many authors have tried to sort out *what* Boltzmann thought and *when* he thought it [11]. For example, he was the first to introduce the use of probabilities into entropy, probably due to reading Maxwell [12]. However, in his writings he uses completely different definitions of probability. Sometimes he used different definitions in the same paper and worse, in the same equation. With this in mind, we attempt to summarize Boltzmann's contribution to entropy with modern notation.

**Thermodynamic Entropy**

In 1871, building on the theory of gases, Boltzmann decided to use the *average* of the energy to represent the relevant information needed for this macroscopic entropy [13],

$$E = \langle H \rangle = \int d\tau_N P_N H \ , \tag{2.1}$$

where $H$ is the Hamiltonian, $d\tau_N = d^{3N}x d^{3N}p$ is the phase space volume element and $P_N$ is the distribution function. At this point he suggests that $P_N$ might be similar to the velocity distribution and writes,

$$P_N = \frac{\exp(-\beta H)}{Z} \ , \tag{2.2}$$

where $Z = \int d\tau_N e^{-\beta H}$, $\beta = 1/T$ (the constant $k$ had not be determined yet). This leads to

$$E = \frac{3}{2}NkT + \langle U \rangle \ , \tag{2.3}$$

where $U$ is the potential in the Hamiltonian.

Connecting this to Clausius' thermodynamic entropy [14] is now difficult since work, $W$, is based on a *microscopic* theory (kinetic theory) while heat is thought of as a macroscopic quantity. The key is that any energy exchange, $\delta E$, needs to be separated into two components, heat and work. To do this we differentiate (2.1),

$$\delta E = \int d\tau_N P_N \delta H + \int d\tau_N H \delta P_N ,\qquad(2.4)$$

where the second integral can be written as $\langle \delta H \rangle = \langle \delta U \rangle$ and comes solely from changes in the potential. This is identified as work, since the change in potential energy is work, $\delta W$. Following energy conservation, it must be then that the second integral is heat, $\delta Q$. Therefore,

$$\delta E = \delta Q + \langle \delta U \rangle .\qquad(2.5)$$

Substituting (2.3) into (2.5) yields,

$$\delta Q = \frac{3}{2} N k \delta T + \delta \langle U \rangle - \langle \delta U \rangle .\qquad(2.6)$$

Dividing both sides by $T$ yields,

$$\frac{\delta Q}{T} = \left[\frac{3}{2} N k \delta T + \delta \langle U \rangle - \langle \delta U \rangle\right] / T .\qquad(2.7)$$

This implies that the right side should be a form for the thermodynamic entropy, $S$. In fact, (2.7) can be rewritten as

$$S = \frac{E}{T} + \log Z + C ,\qquad(2.8)$$

where $C$ is a constant. This is the modern form for thermodynamic entropy.

**Statistical Mechanics**

Through his work above, Boltzmann now has a clear understanding that the arguments for a mechanical theory of heat rest not in the specific mechanics of the particles themselves, but in looking at the probabilities representing the energetic states of the particles. In 1877 he considers "an unrealizable fiction" [15]. He envisions that the phase space of a single particle can be divided into $m$ cells, each with energy $\varepsilon_i$, where $i = 1, 2 \ldots m$. The $i$ indicates a particular *microstate*. He further imagines that there are $N$ particles in this space with $n_i$ particles occupying the $i^{th}$ cell. He then assumes that each cell, has an 'equal chance' of being occupied by a particular particle.[1] He then went on to propose that the distribution of particles for a particular *macrostate* is proportional to the multiplicity for that macrostate,

$$W = \frac{N!}{\prod_{i=1}^{m} n_i!} . \tag{2.9}$$

If $n_i$ is very large, we may apply Stirling's approximation which yields

$$\log W = N \log N - N - \sum_{i=1}^{m} (n_i \log n_i - n_i) . \tag{2.10}$$

Since we know that $\sum_{i=1}^{m} n_i = N$, we can rewrite this as

$$\log W = N \log N - \sum_{i=1}^{m} n_i \log n_i , \tag{2.11}$$

---

[1] Although Boltzmann did not use the principle of insufficient reason, we can make the claim that his argument for 'equally likely' has to do with lack of information which sounds a lot like PIR.

and finally,

$$\log W = -N \sum_{i=1}^{m} \frac{n_i}{N} \log \frac{n_i}{N} . \qquad (2.12)$$

Boltzmann's entropy is written as,

$$\log W = -N \sum_{i=1}^{m} f_i \log f_i , \qquad (2.13)$$

where $f_i = n_i/N$.

Next Boltzmann maximized the log of his multiplicity by using the calculus of variations (CoV). He attempted to find the function, $f$, that maximized $\log W$, given what he knew about the system. For this exercise, he assumed he had a closed system. Therefore, the total number of particles, $N$, and the total energy of the system, $E$, was fixed. We write these constraints as

$$\sum_{i=1}^{m} n_i = N , \qquad (2.14)$$

and

$$\sum_{i=1}^{m} n_i \epsilon_i = E . \qquad (2.15)$$

By CoV (or as Gibbs called it, the "variational principle"), we will now vary $f_i$ in (2.13) and include the constraint information by way of Lagrange multipliers $(\alpha, \beta)$,.so that $\log W$ is maximized,

$$\log W = \log W - \alpha \left( \sum_{i=1}^{m} n_i - N \right) - \beta \left( \sum_{i=1}^{m} n_i \epsilon_i - E \right) .$$

For convenience, we divide by $N$ so that we have

$$\frac{1}{N}\log W = \frac{1}{N}\log W - \alpha\left(\sum_{i=1}^{m} f_i - 1\right) - \beta\left(\sum_{i=1}^{m} f_i\epsilon_i - \frac{E}{N}\right), \qquad (2.16)$$

$$0 = \delta\frac{1}{N}\log W = \delta\left[-\sum_{i=1}^{m} f_i \log f_i - \alpha\left(\sum_{i=1}^{m} f_i - 1\right) - \beta\left(\sum_{i=1}^{m} f_i\epsilon_i - \frac{E}{N}\right)\right], \qquad (2.17)$$

$$0 = -\sum_{i=1}^{m}\log(f_i)\delta f_i - \sum_{i=1}^{m} f_i\frac{1}{f_i}\delta f_i - \alpha\sum_{i=1}^{m}\delta f_i - \beta\sum_{i=1}^{m}\epsilon_i\delta f_i, \qquad (2.18)$$

so that we are left with,

$$0 = \sum_{i=1}^{m}\left(-\log f_i - 1 - \alpha - \beta\epsilon_i\right)\delta f_i, \qquad (2.19)$$

where the inside of the summation must be zero. Therefore,

$$f_i = e^{-1-\alpha}e^{-\beta\epsilon_i}, \qquad (2.20)$$

and using (2.14) we have our final result,

$$f_i = \frac{1}{Z}e^{-\beta\epsilon_i}, \qquad (2.21)$$

where $Z = \sum_{i=1}^{m} e^{-\beta\epsilon_i}$ and $\beta$ is determined by (2.15). This is Boltzmann's version of the *canonical distribution* where (2.21) is the 'probability' of one particle being in the $i^{th}$ microstate.

Additionally, for a gas, we have his final form,

$$\log W = -N\int d\tau_1 f(x,p) \log f(x,p), \qquad (2.22)$$

where $d\tau_1 = d^3x d^3p$.

One can see an immediate problem with this frequentist view; what if the particles interact? Also, there is no mention of irreversibility. The Hamiltonian that Boltzmann uses is definitely time reversible. Where does the second law come into his definition? Boltzmann clearly was the person that first abandons a mechanical argument to describe entropy, but these problems would haunt Boltzmann to the end of his tragic days. For a detailed description see [16]. As a side note, it was Max Plank who first wrote $S = k \log \Omega$, where $\Omega = W$ and established the constant $k$ [17].

## 2.2  Josiah Willard Gibbs (1839–1903)

Intermingled in Boltzmann's accomplishments were Gibbs' revelations. One thing Gibbs realized was that Boltzmann's distribution failed to take into account interactions between particles. Gibbs imagined an *ensemble* of all the possible configurations of a system of $N$ particles with each configuration being a potential microstate, $i$, of the entire system. The macrostate of the system is characterized by a probability distribution of an ensemble. Gibbs' stated that the preferred macrostate is the one that *maximizes* the entropy given some *constraints*. In fact, he made the bold prediction that entropy does not just "tend" to increase, as Clausius stated, but *must* increase. Thus, the entropy that will allow for interacting particles is,

$$S[f] = -\sum_{i=1}^{m} f_i \log f_i \; , \qquad (2.23)$$

where $f_i$ is some function that represents the probability for the system to be in a certain energy state, $\varepsilon_i$ and $i$ represents an $N$ particle microstate, unlike (2.13) This is Gibbs' Entropy [18].

Furthermore, the correct entropy for a gas that will allow for interacting particles is,

$$S[f] = -\int d\tau f(x,p) \log f(x,p) , \qquad (2.24)$$

where $d\tau = d^{3N}x d^{3N}p$ and $f(x,p)$ is some function that represents a probability distribution for a particle including all possible interactions.

Gibbs realized that we can seldom get the exact total energy of a closed system; but the *expectation* of the total energy seems reasonable since we can measure temperature quite accurately. Then (2.14) and (2.15) would be written as,

$$\sum_{i=1}^{m} f_i = 1$$

$$\sum_{i=1}^{m} f_i E_i = \langle E \rangle , \qquad (2.25)$$

where $E_i$ represents the energy of the $i^{th}$ state and we would write (2.21) as,

$$f_i = \frac{1}{Z} e^{-\beta E_i} , \qquad (2.26)$$

where $Z = \sum_{i=1}^{m} e^{-\beta E_i}$ and $\beta$ is determined by (2.25) and the thermodynamic identity so that $\beta = 1/T$. We refer to (2.26) as the more familiar version of the *canonical ensemble*. Instead of a set of fixed, equal energy states, $\varepsilon_i$, we have a variable, $E_i$, that represents the energy of the $i^{th}$ state. Note that this '$i$' is a different one than those in (2.21). Additionally, one need not have a closed system. Since we are dealing with the expected energy, it can fluctuate and even exchange energy with another system. Also, by adding another constraint, we can exchange particles, $N_i$, as well,

$$\sum_{i=1}^{m} f_i N_i = \langle N_i \rangle \;, \tag{2.27}$$

so that,

$$f_i = \frac{1}{Z} e^{-\beta E_i - \gamma N_i} \;, \tag{2.28}$$

where $\gamma$ is determined by (2.27) and the thermodynamic identity so that $\gamma = -\mu/T$, where $\mu$ is the *chemical potential*. We refer to (2.28) as the *grand canonical ensemble*. We hope it is clear that although we continued to use the notation from Boltzmann, $f_i$, it is not, in general, a frequency. Although we did not need a frequency in our derivation, Gibbs thought of the function, $f_i$ as a frequency that represented the probability distribution of an imaginary "ensemble". To see a detailed explanation of why Gibbs' entropy is the correct entropy that actually leads to Clausius' entropy, read the brilliant paper by Jaynes [19].

Finally, we end this section by emphasizing two points from Gibbs' work: 1) We do not have to *assume* that (2.23) or (2.24) only hold when the system is at equilibrium, only that entropy is must increase [20] or that we must maximize the entropy given a set of constraints. 2) The distribution that we get in the end is based on our information, not what *is* or *is not true*. Meaning that they are guesses. But, they are our *best* guesses. If they agree with experiment, then we have all of the relevant information for *that* experiment. If they do not agree with experiment, we have not failed but discovered 'something' new. Therefore, entropy can be used as a tool to learn.

## 2.3 Summary

In this chapter, we made no mention of ergodic theory or so called 'coarse graining'. There was no need. Boltzmann tried to get around using the principle of insufficient reason by using his ergodic hypothesis. Unfortunately it is truly "an unrealizable fiction" that can never be proved by its own definition. Gibbs was and still is criticized for failing to mention in his book how his theory meshes with ergodic hypothesis. Although his health was failing rapidly at the time, he did not *forget* to mention it. Additionally, it must be noted that we do not 'create' a view of the system by an arbitrary or artificial coarse graining. Entropy does not need PIR, ergodic theory or so called coarse graining, it only needs the information that one can measure. Our inference about the system is determined by how much information we have at hand. If we had all information regarding the system we could determine energy states, *exactly* (at least classically and perhaps in the future, quantum mechanically as well).

Last, we rephrase the last comment in the section on Gibbs: We now have tools to learn 'something'. Entropy is a tool for learning. In the cases above we are usually discussing gases. However, that is only because it is information regarding gases that we have examined. In the next chapter we will start to show that the concept of entropy can be used to learn much more.

# Chapter 3

# Information entropy

In the first chapter we focused on probabilities. However, we did not discus how one deals with information in the form of constraints such as an expected value. In the second chapter, we saw that the concept of entropy was used for determining probabilities related to gases by way of constraint information. In this chapter, these ideas start to come together to form a cohesive, general methodology.

## 3.1 Claude Elwood Shannon (1916–2001)

Claude Shannon worked on radar for the U.S. government during WWII. Afterwards, he was interested in the most efficient way to encode a message and how much of it would be received. What he came up with was what he called his 'uncertainty' about the signal [21]. He approached this problem in an interesting way. Instead of creating a mathematical formalism, then trying to *explain* how it applies to his situation, he described what a method should be able to do, then created a mathematical formalism to represent it.

Let $i$ be a set of *mutually exclusive* and *exhaustive* alternatives, such as energy states. Yet the state of the system is unknown. We assign a probability for this state, $p_i$, given some initial (yet insufficient) information, $I$, so that $p_i = p(i|I)$. It would seem reasonable that we would be more uncertain if the probability distribution was broadly peaked, than if it was sharply peaked. How do we determine the amount of information that we need in order to get a sharp peak? Could we define a *measure* that would be consistent. Would it be unique?

Consider a discrete set of $m$, mutually exclusive and exhaustive discrete alternatives, $i = 1, 2 \ldots m$, each with probability $p_i$. Shannon claimed that any measure, $S$, of the amount of information that is *missing* when all we know is a probability distribution must satisfy three axioms:

**Axiom 1** $S$ is a real and continuous function of the probabilities $p_i$, $S[p] = S(p_1, \ldots p_m)$.

**Axiom 2** If all $p_i$s are equal, then $p_i = 1/m$. Thus, $S = S(1/m, \ldots, 1/m) = F(m)$, where $F(m)$ is an monotonic increasing function of $m$.

The third axiom is a consistency requirement. We want the function, $S[p]$, to measure the amount of additional information needed after $I$, to determine the state of the system. Let us assume that the additional information could come in pieces or segments. The consistency requirement is that the order in which these segments come should not matter. Therefore, let us divide the number of states, $m$, into $M$ groups, so that $g = 1, \ldots, M$ and that the probability for the system to be found in a particular group, $g$, is

$$P_g = \sum_{i \in g} p_i \ . \tag{3.1}$$

**Axiom 3** For all possible groupings $g = 1, \ldots, M$ of the states $i = 1, \ldots, m$ we must have

$$S = S_G + \sum_g P_g S_g \ . \tag{3.2}$$

where $S_G$ is the amount of information needed to single out a particular group and the summation is the expected amount of information needed to decide on the actual $i$ within the selected group $g$. This is called the "grouping" property.

For the sake of brevity, we omit the algebraic manipulation to attain Shannon's final result. A brief derivation is given in the appendix of [22]. The unique function that Shannon obtains, as a result of his axioms, to measure the missing amount of information is

$$S[p] = -k \sum_{i=1}^{m} p_i \log p_i \ , \tag{3.3}$$

where $k$ is a constant. John von Neumann examined Shannon's results and told him that uncertainty was a well defined quantity in physics and he should not call it uncertainty. His suggestion was to call it 'entropy', "since no one really knows what entropy is anyway".[1] For a detailed analysis of the applications of his theory, see [23].

---

[1] Shannon's work used binary code, so he further defined his entropy as using a base 2 log, so that log2=1 bit of information. Although it was one of Shannon's colleagues, John Wilder Tukey that coined the term 'bit', Shannon was the first to publish it.

## 3.2 Edwin Thompson Jaynes (1922–1998)

Now is the time when most of what we have been discussing comes to a head. In 1952, Brillouin noticed that it was not coincidental that Shannon's information entropy looked very similar to Gibbs' entropy [24]. Agreeing with this, E. T. Jaynes wrote that Shannon's entropy is more than simply a device for measuring the uncertainty in signals. He proclaimed that it was a generalization of all entropy concepts. Furthermore, since Shannon's entropy was derived by starting with logical assumptions (axioms), as opposed to developing a mathematical formalism first, entropy should be used *a priori*. In other words, when trying to learn something new (new probabilities), we start with entropy and proceed from there. Following Jeffreys' definition of probabilities [25] (and therefore Laplace's) that were ratified by Cox, and using Gibbs' variational principle, he suggested the following:

Let $x$ be a set of discrete values, $x = \{x_1 \ldots x_m\}$ and $p_i$ be a corresponding probability. We wish to know what the probability $p_i$ is, given that all we know is the consistency requirement that all probabilities pertaining to $x$ sum to one,

$$\sum_i^m p_i = 1 \ . \tag{3.4}$$

Using the variational principle, we can write the following,

$$S[p] = S[p] - \alpha \left( \sum_{i=1}^m p_i - 1 \right) \ , \tag{3.5}$$

which, by maximizing (3.3), leads to

$$p_i = \frac{1}{m} \ . \tag{3.6}$$

This is the principle of insufficient reason. We can extend this further to include additional information that constrains what $p_i$ could be, such as the expectation of some function of $x$,

$$\langle f(x) \rangle = \sum_i^m p_i \, f(x_i) \ . \tag{3.7}$$

Using the variational principle again we get,

$$p_i = \frac{1}{Z} e^{-\beta f(x_i)} \ , \tag{3.8}$$

where $Z = \sum_{i=1}^m e^{-\beta f(x_i)}$ and $\beta$ is determined by (3.7),

$$\frac{-\partial \log Z}{\partial \beta} = \langle f(x) \rangle \ . \tag{3.9}$$

Jaynes showed that this is exactly the procedure to attain the canonical (2.26) and grand canonical (2.28) distributions. This method, called MaxEnt (for maximum entropy), states that one should choose a probability that maximizes the Shannon entropy given information in the form of constraints. Thus, Gibbs' entropy is a special case of this new methodology. It is critical to point out that the MaxEnt procedure effectively infers $p_i$ on the basis of some information. If we do not incorporate all the relevant information into the procedure (we have *incomplete* information), then the probability that we attain is a function which reflects this partial knowledge.

However, why choose the probability that maximizes the entropy? In Gibbs' case, an equilibrium argument could be used. However, what about cases of general inference? The answer is that if entropy is thought of as a measure of information, as Shannon described at times, then one should choose the probability that includes the *least* amount of information, just the information given. In other words, other probabilities could be chosen that satisfy the entropy and constraint information, but the probability that maximizes the entropy is the least informative or as some say, the most *honest*.

## 3.3 Relative Entropy

In 1951 Solomon Kullback (1903–1994) introduced an information measure [26] similar to Shannon's,

$$K[p] = \sum_{i=1}^{m} p(x)_i \log \frac{p(x_i)}{q(x_i)} \;, \tag{3.10}$$

where $K[p]$ reflects the 'distance' (although not a distance in the true metric sense) between two probability distributions, $p_i$ and $q_i$. Additionally, we can examine the continuous case,

$$K[p] = \int dx \; p(x) \log \frac{p(x)}{q(x)} \;. \tag{3.11}$$

Here it should be noted that one cannot simply write,

$$S[p] = -\int dx \; p(x) \log p(x) \;, \tag{3.12}$$

for the continuous Shannon case because it is not invariant under coordinate transformation. Jaynes later extended Shannon's entropy to the continuous case,

$$S[p] = -\int dx\ p(x) \log \frac{p(x)}{m(x)}\ ,\qquad(3.13)$$

where $x$ is a continuous variable and $m(x)$ is an "invariant measure"[2] function, proportional to the limiting density of discrete points [27]. Thus, $m(x)$ is necessary for $S[p]$ to remain invariant under coordinate transformations. With this mode of logic, we can generalize Shannon's entropy for the discrete case where,

$$S[p] = -\sum_{i=1}^{m} p(x)_i \log \frac{p(x_i)}{m(x_i)}\ .\qquad(3.14)$$

Where $m(x_i)$ would be a constant for the discrete case,

$$S[p] = -\sum_{i=1}^{m} p(x)_i \log p(x_i) + const\ ,\qquad(3.15)$$

and the constant can be dropped as it does not affect the ranking of the entropy.

## 3.4 Summary

Shannon's entropy started to be used in many aspects of science. Unfortunately, like most good things, it was abused. The old confusion about entropy in thermodynamics resurfaced at this time. What was entropy? Was it what Clausius suggested? How about what Boltzmann or Gibbs suggested? The easy explanation was that they are simply different entropies. However, if they are different, are they related? The solution to this confusion is that they **are** different because they are used for *different* purposes. However, the *process* or method, as outlined by Jaynes, is the same for *each* solution. In other words, we use the same process but different information to arrive at the various 'entropies'. Thus, they are related because the process is the same; entropy is a learning tool, nothing more.

---

[2] The true invariant measure would be the density, $dx\ m(x)$.

It just so happened that at almost the same time, Bayesian statistics received an enormous boost. Since the 1950s, with the works of Jeffreys, Cox, Ramsey [28], Savage [29], DeFinetti [30] and Wald [31], Bayesian probabilities were shown to be superior to frequentist approaches. The real problem was that they were difficult to compute. However, with the major increase in computing power in the 1980s and the rediscovery of Markov Chain Monte Carlo methods in the 1990s, Bayesian methodology exploded and interest in using information entropy to assign probabilities waned.

Still others have offered interesting new perspectives regarding entropy recently. For example, in the lattice theory approach, entropy is a measure on the lattice of questions, which may give rise to a maximum entropy principle [32].

# Part 2: Universal Inference

In 1980, Shore and Johnson realized that one could axiomatize the updating method itself, instead of just the information measure as Shannon did [33]. The method they developed was in contrast with the older MaxEnt method [22], which was designed to assign rather than update probabilities. This was improved upon by others such as John Skilling [34] and Ariel Caticha [35]. The method that came from these improvements was called the method of Maximum (relative) Entropy (ME) and was designed for *updating* probabilities when new information is given in the form of a constraint on the family of allowed posteriors. Thus, ME is a tool for *learning*, like Bayes' Theorem. However, the method was still limited to using information in the form of constraints, unlike Bayes which was limited to information in the form of data.

The major achievement that we will show in the following chapters is that ME is capable of processing information in the form of constraints, like MaxEnt and information in the form of data, as in Bayes' Theorem. Better still is the fact that ME can process both forms *simultaneously*, which Bayes and MaxEnt cannot do alone. Our objective in this part is to define the ME method and illustrate the use of ME by way of proof and example. It is perhaps worth emphasizing that in this approach, entropy is a tool for reasoning which requires no interpretation in terms of heat, multiplicities, disorder, uncertainty, or amount of information.

# Chapter 4

# Entropy as a tool for updating probabilities

The core of this chapter has been previously published in [36].

Our objective is to devise a general method to update from a prior distribution $q(x)$ to a posterior distribution $p(x)$ when new information becomes available. To carry out the update we proceed by ranking the allowed probability distributions according to increasing *preference*. This immediately raises two questions: (a) how is the ranking implemented and (b) what makes one distribution preferable over another? The answer to (a) is that any useful ranking scheme must be transitive (if $P_1$ is better than $P_2$, and $P_2$ is better than $P_3$, then $P_1$ is better than $P_3$), and therefore it can be implemented by assigning a real number $S[P]$ to each $P$ in such a way that if $P_1$ is preferred over $P_2$, then $S[P_1] > S[P_2]$. The preferred $P$ is that which maximizes the "entropy" $S[P]$. This explains why entropies are real numbers and why they are meant to be maximized.

Question (b), the criterion for preference, is implicitly answered once the functional form of the entropy $S[P]$ that defines the ranking scheme is chosen. The basic strategy is inductive. We follow Skilling's method of induction [34]: (1) If an entropy $S[P]$ of universal applicability exists, it must apply to special examples. (2) If in a certain example the best distribution is known, then this knowledge constrains the form of $S[P]$. Finally, (3) if enough examples are known, then $S[P]$ will be completely determined. (Of course, the known examples might turn out to be incompatible with each other, in which case there is no universal $S[P]$ that accommodates them all.)

The known special examples, which are called the "axioms" of ME, reflect the conviction that what was learned in the past is important and should not be easily ignored. The chosen posterior distribution should coincide with the prior as closely as possible and one should only update those aspects of one's beliefs for which corrective new evidence has been supplied. The first two axioms are listed below. (Detailed proofs for the first two axioms are found in [35].)

**Axiom 1: Locality:** *Local information has local effects.*

When the new information does not refer to a domain $D$ of the variable $x$ the conditional probabilities $p(x|D)$ need not be revised. The consequence of the axiom is that non-overlapping domains of $x$ contribute additively to the entropy:

$$S[P] = \int dx\, F(P(x), x) \tag{4.1}$$

where $F$ is some unknown function.

**Axiom 2: Coordinate invariance:** *The ranking should not depend on the system of coordinates.*

The coordinates that label the points $x$ are arbitrary; they carry no information. The consequence of this axiom is that $S[P] = \int dx\, m(x)\Phi(P(x)/m(x))$ involves coordinate invariants such as $dx\, m(x)$ and $P(x)/m(x)$, where the functions $m(x)$ (which is a density) and $\Phi$ are, at this point, still undetermined.

Next we make a second use of the locality axiom and allow domain $D$ to extend over the whole space. Axiom 1 then asserts that *when there is no new information there is no reason to change one's mind.* When there are no constraints the selected posterior distribution should coincide with the prior distribution. This eliminates the arbitrariness in the density $m(x)$: up to normalization $m(x)$ is the prior distribution, $m(x) \propto q(x)$.

In [35] the remaining unknown function $\Phi$ was determined using the following axiom:

**Old Axiom 3: Subsystem independence:** *When a system is composed of subsystems that are **believed** to be independent it should not matter whether the inference procedure treats them separately or jointly.*

Let us be very explicit about what this axiom means. Consider a system composed of two subsystems which our prior evidence has led us to believe are independent. This belief is reflected in the prior distribution: if the subsystem priors are $q_1(x_1)$ and $q_2(x_2)$, then the prior for the whole system is the product $q_1(x_1)q_2(x_2)$. Furthermore suppose that new information is acquired such that $q_1(x_1)$ is updated to $p_1(x_1)$ and that $q_2(x_2)$ is updated to $p_2(x_2)$. Nothing in this new information requires us to revise our previous assessment of independence, therefore there is no need to change our minds, and the function $\Phi$ must be such that the prior for the whole system $q_1(x_1)q_2(x_2)$ should be updated to $p_1(x_1)p_2(x_2)$.

This idea is implemented as follows: First we treat the two subsystems separately. Suppose that for subsystem 1 maximizing

$$S_1[P_1, q_1] = \int dx_1 \, q_1(x_1) \Phi\left(\frac{P_1(x_1)}{q_1(x_1)}\right) , \qquad (4.2)$$

subject to constraints $\mathcal{C}_1$ on the marginal distribution $P_1(x_1) = \int dx_2 \, P(x_1, x_2)$ selects the posterior $p_1(x_1)$. The constraints $\mathcal{C}_1$ could, for example, include normalization, or they could involve the known expected value of a function $f_1(x_1)$,

$$\int dx_1 f_1(x_1) P_1(x_1) = \int dx_1 dx_2 \, f_1(x_1) P(x_1, x_2) = \langle f_1(x_1) \rangle = F_1 . \qquad (4.3)$$

Similarly, suppose that for subsystem 2 maximizing the corresponding $S_2[P_2, q_2]$ subject to constraints $\mathcal{C}_2$ on $P_2(x_2) = \int dx_1 \, P(x_1, x_2)$ selects the posterior $p_2(x_2)$.

Next we treat the subsystems jointly and maximize the joint entropy,

$$S[P, q_1 q_2] = \int dx_1 dx_2 \, q_1(x_1) q_2(x_2) \Phi\left(\frac{P(x_1, x_2)}{q_1(x_1) q_2(x_2)}\right) , \qquad (4.4)$$

subject to the *precisely the same constraints* on the joint distribution $P$. The function $\Phi$ is determined by the requirement that the selected posterior be $p_1 p_2$. As shown in

[35] this leads to the logarithmic form

$$S[P, q] = -\int dx\, P(x) \log \frac{P(x)}{q(x)} \ . \tag{4.5}$$

Next we replace the old axiom 3 by an axiom which is more convincing because it is an explicit requirement of consistency.

**New Axiom 3: Consistency for independent subsystems:** *When a system is composed of subsystems that are **known** to be independent it should not matter whether the inference procedure treats them separately or jointly.*

Again, we have to be very explicit about what this axiom means and how it differs from the old one. When the subsystems are treated separately the inference proceeds exactly as described before: for subsystem 1 maximize the entropy $S_1[P_1, q_1]$ subject to the constraints $\mathcal{C}_1$ to select a posterior $p_1$ and similarly for subsystem 2 to select $p_2$. The important difference is introduced when the subsystems are treated jointly. Since we are only concerned with those special examples where we *know* that the subsystems are independent, we are *required* to search for the posterior within the restricted family of joint distributions that take the form of a product $P = P_1 P_2$; this is an *additional* constraint over and above the original $\mathcal{C}_1$ and $\mathcal{C}_2$.

In the previous case we chose $\Phi$ so as to maintain independence because there was no evidence against it. Here we impose independence by hand as an additional constraint for the stronger reason that the subsystems are known to be independent. At first sight it appears that the new axiom does not place as stringent a restriction on the general form of $\Phi$: it would seem that $\Phi$ has been relieved of its responsibility of enforcing independence because it is up to us to impose it explicitly by hand. However, as we shall see, the fact that we seek an entropy $S$ of *general* applicability and that we require consistency for *all possible* independent subsystems is sufficiently restrictive.

The new constraint $P = P_1 P_2$ is easily implemented by direct substitution. Instead of maximizing the joint entropy, $S[P, q_1 q_2]$, we now maximize

$$S[P_1 P_2, q_1 q_2] = \int dx_1 dx_2 \, q_1(x_1) q_2(x_2) \Phi\left(\frac{P_1(x_1) P_2(x_2)}{q_1(x_1) q_2(x_2)}\right), \tag{4.6}$$

under independent variations $\delta P_1$ and $\delta P_2$ subject to the same constraints $\mathcal{C}_1$ and $\mathcal{C}_2$ and we choose $\Phi$ by imposing that the updating leads to the posterior $p_1(x_1) p_2(x_2)$.

## 4.1 Consistency for identical independent subsystems

Here we show that applying the axiom to subsystems that happen to be identical restricts the entropy functional to a member of the one-parameter family given by

$$S_\eta[P, q] = -\int dx \, P(x) \left(\frac{P(x)}{q(x)}\right)^\eta \quad \text{for} \quad \eta \neq -1, 0 \,. \tag{4.7}$$

Since entropies that differ by additive or multiplicative constants are equivalent in that they induce the same ranking scheme, we could equally well have written

$$S_\eta[P, q] = \frac{1}{\eta(\eta + 1)} \left(1 - \int dx \, P^{\eta+1} q^{-\eta}\right) \,. \tag{4.8}$$

This is convenient because the entropies for $\eta = 0$ and $\eta = -1$ can be obtained by taking the appropriate limits. For $\eta \to 0$ use $y^\eta = \exp \eta \log y \approx 1 + \eta \log y$ to obtain the usual logarithmic entropy, $S_0[P, q] = S[P, q]$ in (4.5). Similarly, for $\eta \to -1$ we get $S_{-1}[P, q] = S[q, P]$.

The proof below is based upon and extends a previous proof by Karbelkar [37]. He showed that belonging to the family of $\eta$-entropies is a sufficient condition to satisfy the consistency axiom for identical systems and he conjectured but did not prove that this was also a necessary condition. Although necessity was not essential to his argument it is crucial for ours because we are seeking the universal updating method. We show below that for identical subsystems there are no acceptable entropies outside this family.

**Proof** First we treat the subsystems separately. For subsystem 1 we maximize the entropy $S_1[P_1, q_1]$ subject to normalization and the constraint $\mathcal{C}_1$ in (4.3). Introduce Lagrange multipliers $\alpha_1$ and $\lambda_1$,

$$\delta \left[ S_1[P_1, q_1] - \lambda_1 \left( \int dx_1 f_1 P_1 - F_1 \right) - \alpha_1 \left( \int dx_1 \, P_1 - 1 \right) \right] = 0, \quad (4.9)$$

which gives

$$\Phi' \left( \frac{P_1(x_1)}{q_1(x_1)} \right) = \lambda_1 f_1(x_1) + \alpha_1 \,, \quad (4.10)$$

where the prime indicates a derivative with respect to the argument, $\Phi'(y) = d\Phi(y)/dy$. For subsystem 2 we need only consider the extreme situation where the constraints $\mathcal{C}_2$ determine the posterior completely: $P_2(x_2) = p_2(x_2)$.

Next we treat the subsystems jointly. The constraints $\mathcal{C}_2$ are easily implemented by direct substitution and thus, we maximize the entropy $S[P_1p_2, q_1q_2]$ by varying over $P_1$ subject to normalization and the constraint $\mathcal{C}_1$ in (4.3). Introduce Lagrange multipliers $\alpha$ and $\lambda$,

$$\delta \left[ S[P_1p_2, q_1q_2] - \lambda \left( \int dx_1 f_1 P_1 - F_1 \right) - \alpha \left( \int dx_1 P_1 - 1 \right) \right] = 0, \quad (4.11)$$

which gives

$$\int dx_2 \, p_2 \Phi' \left( \frac{P_1 p_2}{q_1 q_2} \right) = \lambda[p_2, q_2] f_1(x_1) + \alpha[p_2, q_2] \,, \quad (4.12)$$

where the multipliers $\lambda$ and $\alpha$ are independent of $x_1$ but could in principle be functionals of $p_2$ and $q_2$.

The consistency condition that constrains the form of $\Phi$ is that if the solution to (4.10) is $p_1(x_1)$ then the solution to (4.12) must also be $p_1(x_1)$, and this must be true irrespective of the choice of $p_2(x_2)$. Let us then consider a small change $p_2 \to p_2 + \delta p_2$ that preserves the normalization of $p_2$. First introduce a Lagrange multiplier $\alpha_2$ and rewrite (4.12) as

$$\int dx_2 \, p_2 \Phi' \left( \frac{p_1 p_2}{q_1 q_2} \right) - \alpha_2 \left[ \int dx_2 \, p_2 - 1 \right] = \lambda[p_2, q_2] f_1(x_1) + \alpha[p_2, q_2] \,, \quad (4.13)$$

where we have replaced $P_1$ by the known solution $p_1$ and thereby effectively transformed (4.10) and (4.12) into an equation for $\Phi$. The $\delta p_2(x_2)$ variation gives,

$$\Phi' \left( \frac{p_1 p_2}{q_1 q_2} \right) + \frac{p_1 p_2}{q_1 q_2} \Phi'' \left( \frac{p_1 p_2}{q_1 q_2} \right) = \frac{\delta \lambda}{\delta p_2} f_1(x_1) + \frac{\delta \alpha}{\delta p_2} + \alpha_2 \,. \quad (4.14)$$

Next use (4.10) to eliminate $f_1(x_1)$,

$$\Phi'\left(\frac{p_1 p_2}{q_1 q_2}\right) + \frac{p_1 p_2}{q_1 q_2}\Phi''\left(\frac{p_1 p_2}{q_1 q_2}\right) = A[p_2, q_2]\Phi'\left(\frac{p_1}{q_1}\right) + B[p_2, q_2], \tag{4.15}$$

where

$$A[p_2, q_2] = \frac{1}{\lambda_1}\frac{\delta\lambda}{\delta p_2} \quad \text{and} \quad B[p_2, q_2] = -\frac{\delta\lambda}{\delta p_2}\frac{\alpha_1}{\lambda_1} + \frac{\delta\alpha}{\delta p_2} + \alpha_2, \tag{4.16}$$

are at this point unknown functionals of $p_2$ and $q_2$. Differentiating (4.15) with respect to $x_1$ the $B$ term drops out and we get

$$A[p_2, q_2] = \left[\frac{d}{dx_1}\Phi'\left(\frac{p_1}{q_1}\right)\right]^{-1}\frac{d}{dx_1}\left[\Phi'\left(\frac{p_1 p_2}{q_1 q_2}\right) + \frac{p_1 p_2}{q_1 q_2}\Phi''\left(\frac{p_1 p_2}{q_1 q_2}\right)\right], \tag{4.17}$$

which shows that $A$ is not a functional of $p_2$ and $q_2$ but a mere function of $p_2/q_2$. Substituting back into (4.15) we see that the same is true for $B$. Therefore (4.15) can be written as

$$\Phi'(y_1 y_2) + y_1 y_2 \Phi''(y_1 y_2) = A(y_2)\Phi'(y_1) + B(y_2), \tag{4.18}$$

where $y_1 = p_1/q_1$, $y_2 = p_2/q_2$, and $A(y_2)$, $B(y_2)$ are unknown functions of $y_2$. If we specialize to *identical* subsystems for which we can exchange the labels $1 \leftrightarrow 2$, we get

$$A(y_2)\Phi'(y_1) + B(y_2) = A(y_1)\Phi'(y_2) + B(y_1). \tag{4.19}$$

To find the unknown functions $A$ and $B$ differentiate with respect to $y_2$,

$$A'(y_2)\Phi'(y_1) + B'(y_2) = A(y_1)\Phi''(y_2) \tag{4.20}$$

and then with respect to $y_1$ to get

$$\frac{A'(y_1)}{\Phi''(y_1)} = \frac{A'(y_2)}{\Phi''(y_2)} = a = \text{const} . \qquad (4.21)$$

Integrating,

$$A(y_1) = a\Phi'(y_1) + b . \qquad (4.22)$$

Substituting $A$ and $B$ back into (4.20) and integrating gives

$$B'(y_2) = b\Phi''(y_2) \quad \text{and} \quad B(y_2) = b\Phi'(y_2) + c , \qquad (4.23)$$

where $b$ and $c$ are constants. We can check that $A(y)$ and $B(y)$ are indeed solutions of (4.19). Substituting into (4.18) gives

$$\Phi'(y_1 y_2) + y_1 y_2 \Phi''(y_1 y_2) = a\Phi'(y_1)\Phi'(y_2) + b\left[\Phi'(y_1) + \Phi'(y_2)\right] + c . \qquad (4.24)$$

This is a peculiar differential equation. We can think of it as one differential equation for $\Phi'(y_1)$ for each given constant value of $y_2$ but there is a complication in that the various (constant) coefficients $\Phi'(y_2)$ are themselves unknown. To solve for $\Phi$, choose a fixed value of $y_2$, say $y_2 = 1$,

$$y\Phi''(y) - \eta\Phi'(y) - \kappa = 0 , \qquad (4.25)$$

where $\eta = a\Phi'(1) + b - 1$ and $\kappa = b\Phi'(1) + c$. To eliminate the constant $\kappa$ differentiate with respect to $y$,

$$y\Phi''' + (1-\eta)\Phi'' = 0 , \tag{4.26}$$

which is a linear homogeneous equation and is easy to integrate. For a generic value of $\eta$ the solution is

$$\Phi''(y) \propto y^{\eta-1} \Rightarrow \Phi'(y) = \alpha y^\eta + \beta . \tag{4.27}$$

The constants $\alpha$ and $\beta$ are chosen so that this is a solution of (4.24) for all values of $y_2$ (and not just for $y_2 = 1$). Substituting into (4.24) and equating the coefficients of various powers of $y_1 y_2$, $y_1$, and $y_2$ gives three conditions on the two constants $\alpha$ and $\beta$,

$$\alpha(1+\eta) = a\alpha^2, \quad 0 = a\alpha\beta + b\alpha, \quad \beta = a\beta^2 + 2b\beta + c . \tag{4.28}$$

The nontrivial ($\alpha \neq 0$) solutions are $\alpha = (1+\eta)/a$ and $\beta = -b/a$, while the third equation gives $c = b(1-b)/4a$. We conclude that for generic values of $\eta$ the solution of (4.24) is

$$\Phi(y) = \frac{1}{a} y^{\eta+1} - \frac{b}{a} y + C , \tag{4.29}$$

where $C$ is a new constant. Choosing $a = -\eta(\eta+1)$ and $b = 1 + Ca$ we obtain (4.8).

For the special values $\eta = 0$ and $\eta = -1$ one can either first take the limit of the differential equation (4.26) and then find the relevant solutions, or one can first solve the differential equation for general $\eta$ and then take the limit of the solution (4.8) as described earlier. Either way one obtains (up to additive and multiplicative constants which have no effect on the ranking scheme) the entropies $S_0[P, q] = S[P, q]$ and $S_{-1}[P, q] = S[q, P]$.

## 4.2 Consistency for non-identical subsystems

Let us summarize our results so far. The goal is to update probabilities by ranking the distributions according to an entropy $S$ that is of general applicability. The functional form of the entropy $S$ has been constrained down to a member of the one-dimensional family $S_\eta$. One might be tempted to conclude (see [37, 4]) that there is no $S$ of universal applicability; that inferences about different systems could be carried out with different $\eta$-entropies. But we have not yet exhausted the full power of our new axiom 3.

To proceed further we ask: What is $\eta$? Is it a property of the individual carrying out the inference or of the system under investigation? The former makes no sense; we insist that the updating must be objective in that different individuals with the same prior and the same information must make the same inference. Therefore, the remaining alternative is that the "inference parameter", $\eta$ must be a characteristic of the system.

Consider two different systems characterized by $\eta_1$ and $\eta_2$. Let us further suppose that these systems are independent (perhaps system 1 is here on Earth while the other lives in a distant galaxy) so that they fall under the jurisdiction of the new axiom 3; inferences about system 1 are carried out with $S_{\eta_1}[P_1, q_1]$ while inferences about system 2 require $S_{\eta_2}[P_2, q_2]$. For the combined system we are also required to use an $\eta$-entropy $S_\eta[P_1 P_2, q_1 q_2]$. The question is what $\eta$ do we choose that will lead to consistent inferences whether we treat the systems separately or jointly. The results of the previous section indicate that a joint inference with $S_\eta[P_1 P_2, q_1 q_2]$ is equivalent to separate inferences with $S_\eta[P_1, q_1]$ and $S_\eta[P_2, q_2]$. Therefore we must choose $\eta = \eta_1$ and also $\eta = \eta_2$ which is possible only when $\eta_1 = \eta_2$. But this is not all: any other system whether here on Earth or elsewhere that happens to be independent of the distant system 2 must also be characterized by the same inference parameter $\eta = \eta_2 = \eta_1$ even if it is correlated with system 1. Thus all systems have the same $\eta$ whether they are independent or not.

The power of a consistency argument resides in its universal applicability: if a general expression for $S[P, q]$ exists then it must be of the form $S_\eta[P, q]$ where $\eta$ is a universal constant. The remaining problem is to determine this universal $\eta$. One possibility is to determine $\eta$ experimentally: are there systems for which inferences based on a known value of $\eta$ have repeatedly led to success? The answer is yes; they are quite common.

The next step in our argument is provided by the work of Jaynes [22] who showed that statistical mechanics and thus thermodynamics are theories of inference based on the value $\eta = 0$. His method, called MaxEnt, can be interpreted as the special case of the ME when one updates from a uniform prior using the Gibbs-Shannon entropy. Thus, it is an experimental fact without any known exceptions that inferences about *all* physical, chemical and biological systems that are in thermal equilibrium or close to it can be carried out by assuming that $\eta = 0$. Let us emphasize that this is not an obscure and rare example of purely academic interest; these systems comprise essentially all of natural science. (Included is every instance where it is useful to introduce a notion of temperature.)

In conclusion: consistency for non-identical systems requires that $\eta$ be a universal constant and there is abundant experimental evidence for its value being $\eta = 0$. Other $\eta$-entropies may be useful for other purposes but the logarithmic entropy $S[P, q]$ in (4.5) provides the only consistent ranking criterion for updating probabilities that can claim general applicability.

## 4.3 Summary

We have shown that Skilling's method of induction has led to a unique general theory of inductive inference, the ME method. The whole approach is extremely conservative. First, the axioms merely instruct us what not to update – do not change your mind except when forced by new information. Second, the validity of the method does not depend on any particular interpretation of the notion of entropy.

We improve on the previous derivation [35] by adopting a considerably weaker axiom that deals with independent subsystems. This axiom is phrased similarly to the one proposed by Shore and Johnson [33]: When two systems are independent it should not matter whether the inference procedure treats them separately or jointly. The merit of such a consistency axiom is that it is very compelling.

Nevertheless, the mathematical implementation of the Shore and Johnson axiom has been criticized by Karbelkar [37] and by Uffink [4]. In their view it fails to single out the usual logarithmic entropy as the unique tool for updating. It merely restricts the form of the entropy to a one-dimensional continuum labeled by a parameter $\eta$. The resulting $\eta$-entropies are equivalent to those proposed by Renyi [38] and by Tsallis [39] in the sense that they update probabilities in the same way.

However, this chapter went beyond the insights of Karlbelkar and Uffink, and showed that our consistency axiom selects a unique, universal value for the parameter $\eta$ and this value ($\eta = 0$) corresponds to the usual logarithmic entropy. The advantage of our approach is that it shows precisely how it is that $\eta$-entropies with $\eta \neq 0$ are ruled out as tools for updating.

# Chapter 5

# Bayes updating

The core of this chapter has been previously published in [36].

Once again, our objective is to devise a general method to update from a prior distribution to a posterior distribution when new information is obtained. We start this section by drawing a distinction between Bayes' *theorem*, which is a straightforward consequence of the product rule for probabilities, and Bayes' *rule*, which is a method to update from a prior distribution to a posterior distribution when we have information in the form of data. Although Bayes' Theorem is used for updating, Cox's derivation makes no mention of updating. It is simply a relationship between conditional probabilities. This distinction is mostly pedagogical yet important.

We show that Bayes' rule can be derived as a special case of the ME method, a result that was first obtained by Williams [42, 43] long before the logical status of the ME method had been sufficiently clarified. The virtue of our derivation, which hinges on translating information in the form of data into constraints that can be processed using ME, is that it is particularly clear. It throws light on Bayes' rule and demonstrates its complete compatibility with ME updating. A slight generalization of the same ideas shows that Jeffreys' updating rule is also a special case of the ME method.

## 5.1 Bayes' theorem and Bayes' rule

The goal here is to update our beliefs about the values of one or several quantities $\theta \in \Theta$ on the basis of observed values of variables $x \in \mathcal{X}$ and of the known relation between them represented by a specific model. The first important point to make is that attention must be focused on the joint distribution $P_{\text{old}}(x, \theta)$. Indeed, being a consequence of the product rule, Bayes' theorem requires that $P_{\text{old}}(x, \theta)$ be defined and that assertions such as "$x$ and $\theta$" be meaningful; the relevant space is neither $\mathcal{X}$ nor $\Theta$ but the product $\mathcal{X} \times \Theta$. The label "old" is important. It has been attached to the joint distribution $P_{\text{old}}(x, \theta)$ because this distribution codifies our beliefs about $x$ and about $\theta$ before the information contained in the actual data has been processed. The standard derivation of Bayes' theorem invokes the product rule,

$$P_{\text{old}}(x, \theta) = P_{\text{old}}(x) P_{\text{old}}(\theta|x) = P_{\text{old}}(\theta) P_{\text{old}}(x|\theta) \;, \tag{5.1}$$

so that

$$P_{\text{old}}(\theta|x) = P_{\text{old}}(\theta) \frac{P_{\text{old}}(x|\theta)}{P_{\text{old}}(x)} \;. \tag{Bayes' theorem}$$

It is important to realize that at this point there has been no updating. Our beliefs have not changed. All we have done is rewrite what we knew all along in $P_{\text{old}}(x, \theta)$. Bayes' *theorem* is an identity that follows from requirements on how we should consis-tently assign degrees of belief. Whether the justification of the product rule is sought through Cox's consistency requirement and regraduation or through a Dutch book betting coherence argument, the theorem is valid irrespective of whatever data will be or has been collected. Our notation, with the label "old" throughout, makes this point explicit.

The real updating from the old prior distribution $P_{\text{old}}(\theta)$ to a new posterior distribution $P_{\text{new}}(\theta)$ occurs when we take into account the values of $x$ that have actually been observed, which we will denote, $x'$. This requires a new assumption and the natural choice is that the updated distribution $P_{\text{new}}(\theta)$ be given by Bayes' *rule*,

$$P_{\text{new}}(\theta) = P_{\text{old}}(\theta|x') \ . \qquad \text{(Bayes rule)}$$

Combining Bayes' theorem with Bayes' rule leads to the standard equation for Bayes updating,

$$P_{\text{new}}(\theta) = P_{\text{old}}(\theta) \frac{P_{\text{old}}(x'|\theta)}{P_{\text{old}}(x')} \ . \qquad (5.2)$$

The assumption embodied in Bayes' rule is extremely reasonable: we maintain those old beliefs about $\theta$ that are consistent with data values that have turned out to be true. Data values that were not observed are discarded because they are now known to be false.

This argument is indeed so compelling that it may seem unnecessary to seek any further justification for the Bayes' rule assumption. However, we deal here with such a basic algorithm for information processing – it is fundamental to all experimental science – that even such a self evident assumption should be carefully examined and its compatibility with the ME method should be verified.

## 5.2 Updating with data using the ME method

In Bayesian inference, it is assumed that one always has a prior probability based on some prior information. When new information is obtained, the old probability (the prior) is *updated* to a new probability (the posterior). If one has no prior information, then one uses an *ignorant*[1] prior [44]. Therefore, our first concern when using the ME method to update from a prior to a posterior distribution is to define the space in which the search for the posterior will be conducted. We wish to infer something about the values of one or several quantities, $\theta \in \Theta$, on the basis of three pieces of information: prior information about $\theta$ (the prior), the known relationship between $x$ and $\theta$ (the model), and the observed values of the data $x \in \mathcal{X}$. Since we are concerned with both $x$ *and* $\theta$, the relevant space is neither $\mathcal{X}$ nor $\Theta$ but the product $\mathcal{X} \times \Theta$ and our attention must be focused on the joint distribution $P(x, \theta)$. The selected joint posterior $P_{\text{new}}(x, \theta)$ is that which maximizes the entropy,

$$S[P, P_{\text{old}}] = -\int dx d\theta \; P(x, \theta) \log \frac{P(x, \theta)}{P_{\text{old}}(x, \theta)} \;, \tag{5.3}$$

---

[1] Ignorant priors are used in common practice. However, term *ignorant* prior is loaded with controversy. What is an "ignorant" prior? One argument is that one always knows some information, such as the likelihood. In which case we may use a Jeffreys prior. Or perhaps a flat prior is appropriate, which we use in examples out of convenience. We do not hope to end this debate here where our focus is on the process of updating.

subject to the appropriate constraints. $P_{\text{old}}(x,\theta)$ contains our prior information which we call the *joint prior*. To be explicit,

$$P_{\text{old}}(x,\theta) = P_{\text{old}}(\theta) P_{\text{old}}(x|\theta) \;, \tag{5.4}$$

where $P_{\text{old}}(\theta)$ is the traditional Bayesian prior and $P_{\text{old}}(x|\theta)$ is the likelihood. It is important to note that they *both* contain prior information. The Bayesian prior is defined as containing prior information. However, the likelihood is not traditionally thought of in terms of prior information. Of course it is reasonable to see it as such because the likelihood represents the model (the relationship between $\theta$ and $x$) that has already been established. Thus we consider both pieces, the Bayesian prior and the likelihood to be *prior* information.

The new information is the *observed data*, $x'$, which in the ME framework must be expressed in the form of a constraint on the allowed posteriors. The family of posteriors that reflects the fact that $x$ is now known to be $x'$ is such that

$$P(x) = \int d\theta \; P(x,\theta) = \delta(x - x') \;. \tag{5.5}$$

This amounts to an *infinite* number of constraints on $P(x,\theta)$: for each value of $x$ there is one constraint and one Lagrange multiplier $\lambda(x)$.

Maximizing (5.3), subject to the constraints (5.5) plus normalization,

$$\delta \left\{ S + \alpha \left[ \int dx d\theta \; P(x,\theta) - 1 \right] + \int dx \; \lambda(x) \left[ \int d\theta \; P(x,\theta) - \delta(x - x') \right] \right\} = 0 \;, \tag{5.6}$$

yields the joint posterior,

$$P_{\text{new}}(x,\theta) = \frac{1}{Z} P_{\text{old}}(x,\theta) \, e^{\lambda(x)} \;, \tag{5.7}$$

where $Z$ is a normalization constant,

$$Z = e^{-\alpha+1} = \int dx d\theta \, P_{\text{old}}(x,\theta) \, e^{\lambda(x)} \;, \tag{5.8}$$

and $\lambda(x)$ is determined from (5.5),

$$\frac{1}{Z}\int d\theta \, P_{\text{old}}(x,\theta) \, e^{\lambda(x)} = \frac{1}{Z} P_{\text{old}}(x) \, e^{\lambda(x)} = \delta(x-x') \;. \tag{5.9}$$

The final expression for the joint posterior is determined by substituting $e^{\lambda(x)}$ back into (5.7),

$$P_{\text{new}}(x,\theta) = \frac{P_{\text{old}}(x,\theta)\,\delta(x-x')}{P_{\text{old}}(x)} = \delta(x-x') P_{\text{old}}(\theta|x) \;, \tag{5.10}$$

and the marginal posterior distribution for $\theta$ is

$$P_{\text{new}}(\theta) = \int dx\, P_{\text{new}}(x,\theta) = P_{\text{old}}(\theta|x') \;, \tag{5.11}$$

which is the familiar Bayes' conditionalization rule.

To summarize: $P_{\text{old}}(x,\theta) = P_{\text{old}}(x) P_{\text{old}}(\theta|x)$ is updated to $P_{\text{new}}(x,\theta) = P_{\text{new}}(x) P_{\text{new}}(\theta|x)$ with $P_{\text{new}}(x) = \delta(x-x')$ fixed by the observed data while $P_{\text{new}}(\theta|x) = P_{\text{old}}(\theta|x)$ remains unchanged. We see that in accordance with the minimal updating philosophy that drives the ME method *one only updates those aspects of one's beliefs for which* *corrective new evidence (in this case, the data) has been supplied.*

**Example – Laplace's Succession Rule**

Here we demonstrate how ME would be used in a problem that is traditionally solved using Bayes Rule: A $k$-sided die has been tossed $n$ times and the $i^{th}$ side comes up $m_i$ times. Given this information, we would like to determine the probability of getting the $i^{th}$ side in the next toss, $\theta_i$. We rewrite (5.3) to reflect our information,

$$S[P, P_{\text{old}}] = -\sum_m \int d\theta \ P(m, \theta|n) \log \frac{P(m, \theta|n)}{P_{\text{old}}(m, \theta|n)} \tag{5.12}$$

where $m = (m_1, \ldots, m_k)$ with $\sum_{i=1}^{k} m_i = n$, and $\theta = (\theta_1, \ldots, \theta_k)$ with $\sum_{i=1}^{k} \theta_i = 1$. We write the constraints (note the use of the Kronecker delta for the discrete case of the die),

$$\sum_m \int d\theta \ P(m, \theta|n) = 1 \ . \tag{5.13}$$

$$P(m|n) = \int d\theta \ P(m, \theta|n) = \delta_{mm'} \ , \tag{5.14}$$

where $m'$ is the actual *observed* quantity. Now we maximize the entropy given the constraints with respect to $P(m, \theta|n)$,

$$\delta \left\{ \begin{array}{l} S + \alpha \left[ \sum_m \int d\theta \ P(m, \theta|n) - 1 \right] \\ + \sum_m \lambda_m \left[ \int d\theta \ P(m, \theta|n) - \delta_{mm'} \right] \end{array} \right\} = 0 \ , \tag{5.15}$$

so that the selected posterior is

$$P_{\text{new}}(m, \theta|n) = \frac{1}{Z} P_{\text{old}}(m, \theta|n) \, e^{\lambda_m} \ , \tag{5.16}$$

where the normalization $Z$, is

$$Z = e^{-\alpha+1} = \sum_m \int d\theta\, P_{\text{old}}(m,\theta|n)\, e^{\lambda_m}\ , \tag{5.17}$$

and the Lagrange multipliers $\lambda_m$ are determined from (5.14),

$$\frac{1}{Z}\int d\theta\, P_{\text{old}}(m,\theta|n)\, e^{\lambda_m} = \frac{1}{Z} P_{\text{old}}(m|n)\, e^{\lambda_m} = \delta_{mm'}\ . \tag{5.18}$$

Therefore, substituting $e^{\lambda_m}$ back into (5.16),

$$P_{\text{new}}(m,\theta|n) = \frac{P_{\text{old}}(m,\theta|n)\,\delta_{mm'}}{P_{\text{old}}(m|n)} = \delta_{mm'} P_{\text{old}}(\theta|m,n)\ . \tag{5.19}$$

The new marginal distribution for $\theta$ is

$$P_{\text{new}}(\theta) = \sum_m d\theta P_{\text{new}}(m,\theta|n) = P_{\text{old}}(\theta|m',n)\ . \tag{5.20}$$

We need to determine $P_{\text{old}}(\theta|m',n)$. We start by using the product rule,

$$P_{\text{old}}(\theta|m',n) = P_{\text{old}}(\theta|n)\frac{P_{\text{old}}(m'|\theta,n)}{P_{\text{old}}(m'|n)}\ . \tag{5.21}$$

This is of course *Bayes' rule*. The marginal likelihood, $P_{\text{old}}(m'|n)$ (also called the evidence) is simply a normalization constant. Dice are modeled in terms of multinomial distributions. The probability that casting a $k$-sided die $n$ times yielding $m_i$ instances

for the $i^{th}$ face is

$$P_{\text{old}}(m|\theta) = P_{\text{old}}(m_1...m_k|\theta_1...\theta_k, n) = \frac{n!}{m_1!...m_k!}\theta_1^{m_1}...\theta_k^{m_k} , \quad (5.22)$$

For our problem, we will let $\theta_i = \theta$, $\sum_{j\neq i}^{k} \theta_j = (1-\theta)$ and $m_i = m$. The particular form of $P_{\text{old}}(\theta|n)$ is not important for our current purpose so for the sake of definiteness we can choose it flat, and in this case, constant. Being a constant, the prior can come out of the integral and cancel with the same constant in the numerator. Now $P_{\text{old}}(\theta|m', n)$ from (5.20) can be written as

$$P_{\text{new}}(\theta) = P_{\text{old}}(\theta|m', n) = \frac{\binom{n}{m'}\theta^{m'}(1-\theta)^{n-m'}}{\int_0^1 \binom{n}{m'}\tilde{\theta}^{m'}(1-\tilde{\theta})^{n-m'}d\tilde{\theta}} = \frac{\theta^{m'}(1-\theta)^{n-m'}}{\int_0^1 \tilde{\theta}^{m'}(1-\tilde{\theta})^{n-m'}d\tilde{\theta}} , \quad (5.23)$$

where the denominator can be solved using a beta function, $B$,

$$B(p, q) = \int_0^1 t^{p-1'}(1-t)^{q-1}dt = \frac{(p-1)!(q-1)!}{(p+q-1)!} \quad (5.24)$$

where we let $p = m + 1$ and $q = n - m + 1$. The result is

$$P_{\text{new}}(\theta) = \frac{(n+1)!}{m'!(n-m')!}\theta^{m'}(1-\theta)^{n-m'} . \quad (5.25)$$

It is important to note that $P_{\text{new}}(\theta)$ is the probability distribution of all possible $\theta$'s. Therefore, we cannot determine without certainty the value of $\theta$. However, we can choose to summarize our uncertain state of knowledge by identifying the $\theta$ that is "most likely" to give us the best results in experiments involving $\theta$. There are traditionally three methods used for this choice: the *mean*, the *mode* and the *median*.

To determine the *mean* value for the parameter, $\theta$ for this problem, we use

$$\langle\theta\rangle = \int_0^1 d\theta\ \theta P_{\text{new}}(\theta)\ . \tag{5.26}$$

For our example, this we have,

$$\langle\theta\rangle = \frac{m+1}{n+2}\ . \tag{5.27}$$

This result is also called the Laplace succession rule.

**Example – Jeffreys' Conditioning Rule**

In this example we consider a slightly more general situation where there is some uncertainty about some data that was collected. In this case, the marginal $P(x)$ in (5.5) is not a $\delta$ function but a known distribution, $P_D(x)$. Therefore, the constraints are

$$P(x) = \int d\theta\ P(x,\theta) = P_D(x)\ . \tag{5.28}$$

The selected posterior is still given by (5.7) with $Z$ given by (5.8) but the multipliers $\lambda(x)$ are now determined from (5.28),

$$\int d\theta\ P_{\text{old}}(x,\theta)\frac{e^{\lambda(x)}}{Z} = P_{\text{old}}(x)\frac{e^{\lambda(x)}}{Z} = P_D(x)\ , \tag{5.29}$$

so that

$$e^{\lambda(x)} = Z\frac{P_D(x)}{P_{\text{old}}(x)}\ . \tag{5.30}$$

Next substitute into (5.7),

$$P_{\text{new}}(x, \theta) = \frac{P_{\text{old}}(x, \theta)}{P_{\text{old}}(x)} P_D(x) = P_{\text{old}}(\theta|x) P_D(x) \tag{5.31}$$

The new marginal distribution for $\theta$ is

$$P_{\text{new}}(\theta) = \int dx P_{\text{new}}(x, \theta) = \int dx\, P_{\text{old}}(\theta|x) P_D(x) \ , \tag{5.32}$$

as expected. This is known as *Jeffrey's rule of conditioning*. It amounts to setting

$$P_{\text{new}}(x, \theta) = P_{\text{new}}(\theta|x) P_{\text{new}}(x) \ , \tag{5.33}$$

with the actual updating being $P_{\text{new}}(\theta|x) = P_{\text{old}}(\theta|x)$ and $P_{\text{new}}(x) = P_D(x)$.

## 5.3 Summary

We explored the compatibility of Bayes and ME updating. After pointing out the distinction between Bayes' theorem and the Bayes' updating rule, we showed that Bayes' rule is a special case of ME updating by translating information in the form of data into constraints that can be processed using ME. This implies that ME is *capable of reproducing every aspect of orthodox Bayesian inference* and proves the complete compatibility of Bayesian and entropy methods.

# Chapter 6

# Universal updating

The core of this chapter has been previously published in [45]

When using Bayes' rule, it is quite common to impose constraints on the prior distribution. In some cases these constraints are also satisfied by the posterior distribution, but these are special cases. In general, constraints imposed on priors do not "propagate" to the posteriors. Although Bayes' rule can handle *some* constraints, we seek a procedure capable of enforcing *any* constraint on the posterior distributions. Thus, in the past, Bayes's Rule has been used when information in the form of data is present.

MaxEnt was used when information in the form of constraints (such as moments) was present and then only to assign probabilities. The ME method extended this method so that one could update from a prior to a posterior when constraint information was obtained. However, the relationship between these two methods has always been controversial, even though it is common practice to use MaxEnt to assign priors.

In the previous chapter we proved that Bayes is a special case of ME and therefore the compatibility of the methods. In this chapter, we do something that to our knowledge has never been done: We update with data **and** constraints *simultaneously*. When using ME to achieve this, we arrive at a final result that is the "canonical" form of the posterior distribution.

This result is deceivingly simple: the likelihood is modified by a "canonical" exponential factor. Although this result is very simple, it should be handled with caution: once we consider several sources of information such as multiple constraints we must confront the problem of non-commutivity. We discuss the question of whether they should be processed simultaneously, or sequentially, and in what order. Our general conclusion is that these different alternatives correspond to different states of information and accordingly we expect that they will lead to different inferences.

To illustrate the differences, a multinomial example of die tosses is solved in some detail for two problems. They appear superficially similar but are in fact very different. The first die problem requires that the constraints be processed sequentially. This corresponds to the familiar situation of using MaxEnt to derive a prior and then using Bayes to process data. The second die problem, which requires that the constraints be processed simultaneously, provides a clear example that lies *beyond* the reach of Bayes' rule or MaxEnt alone.

## 6.1 The Canonical Form

Here we process two forms of information: expected values and data, *simultaneously*. For simplicity we will refer to these expected values as *moments* although they can be considerably more general. The solution resembles Bayes' Rule. In fact, if there are no moment constraints then the method produces Bayes rule *exactly*. If there is no data, then the MaxEnt solution is produced.

We generalize the results of the previous chapter to include additional information about $\theta$ in the form of a constraint on the expected value of some function $f(\theta)$,

$$\int dx d\theta \, P(x,\theta) f(\theta) = \langle f(\theta) \rangle = F \; . \tag{6.1}$$

We emphasize that constraints imposed at the level of the prior need not be satisfied by the posterior. What we do here differs from the standard Bayesian practice in that we *require* the constraint to be satisfied by the posterior distribution.

Maximizing the joint entropy, (5.3), subject to normalization, the data constraint,

$$P(x) = \int d\theta \, P(x,\theta) = \delta(x - x') \; , \tag{6.2}$$

and the moment constraint (6.1),

$$\delta \left\{ \begin{array}{c} S + \alpha \left[ \int dx d\theta \, P(x,\theta) - 1 \right] \\ + \beta \left[ \int dx d\theta \, P(x,\theta) f(\theta) - F \right] \\ + \left[ \int dx \lambda(x) \int d\theta \, P(x,\theta) - \delta(x - x') \right] \end{array} \right\} = 0 \tag{6.3}$$

yields the joint posterior,

$$P_{\text{new}}(x,\theta) = P_{\text{old}}(x,\theta) \, e^{\alpha - 1 + \lambda(x) + \beta f(\theta)} \; . \tag{6.4}$$

The first Lagrange multiplier is determined by substituting (6.4) into the normalization constraint $\int dx d\theta \, P(x,\theta) = 1$ which yields,

$$Z = e^{-\alpha+1} = \int dx d\theta e^{\lambda(x)+\beta f(\theta)} P_{\text{old}}(x,\theta) \tag{6.5}$$

and (6.4) can be rewritten as

$$P_{\text{new}}(x,\theta) = \frac{1}{Z} P_{\text{old}}(x,\theta) \, e^{\lambda(x)+\beta f(\theta)} \, . \tag{6.6}$$

The Lagrange multipliers $\lambda(x)$ are determined by substituting (6.6) into (6.2)

$$e^{\lambda(x)} \int d\theta e^{\beta f(\theta)} P_{\text{old}}(x,\theta) = Z \, \delta(x - x\acute{}) \, , \tag{6.7}$$

so that the joint posterior becomes

$$P_{\text{new}}(x,\theta) = \frac{1}{\zeta(x,\beta)} P_{\text{old}}(x,\theta) \delta(x - x\acute{}) e^{\beta f(\theta)} \, . \tag{6.8}$$

where $\zeta(x,\beta) = \int d\theta e^{\beta f(\theta)} P_{\text{old}}(x,\theta)$.

The Lagrange multiplier $\beta$ is determined by first substituting the posterior into (6.1)

$$\int dx d\theta \left[ \frac{1}{\zeta(x,\beta)} P_{\text{old}}(x,\theta) \delta(x - x\acute{}) e^{\beta f(\theta)} \right] f(\theta) = F \, , \tag{6.9}$$

which can be rewritten as

$$\int dx \left[ \frac{1}{\zeta(x,\beta)} \int d\theta e^{\beta f(\theta)} P_{\text{old}}(x,\theta) f(\theta) \right] \delta(x - x\acute{}) = F \, . \tag{6.10}$$

Integrating over $x$ yields,

$$\frac{\int d\theta e^{\beta f(\theta)} P_{\text{old}}(x',\theta) f(\theta)}{\zeta(x',\beta)} = F \tag{6.11}$$

where $\zeta(x,\beta) \to \zeta'(x',\beta) = \int d\theta e^{\beta f(\theta)} P_{\text{old}}(x',\theta)$. Now $\beta$ can be determined by

$$\frac{\partial \ln \zeta'}{\partial \beta} = F \ . \tag{6.12}$$

The final step is to marginalize the posterior, $P_{\text{new}}(x,\theta)$ over $x$ to get our updated probability,

$$P_{\text{new}}(\theta) = P_{\text{old}}(x',\theta) \frac{e^{\beta f(\theta)}}{\zeta(x',\beta)} \tag{6.13}$$

Additionally, this result can be rewritten using the product rule as

$$P_{\text{new}}(\theta) = P_{\text{old}}(\theta) P_{\text{old}}(x'|\theta) \frac{e^{\beta f(\theta)}}{\zeta'} \ , \tag{6.14}$$

where $\zeta'(x',\beta) = \int d\theta e^{\beta f(\theta)} P_{\text{old}}(\theta) P_{\text{old}}(x'|\theta)$. The right side resembles Bayes rule, where the term $P_{\text{old}}(x'|\theta)$ is the standard Bayesian likelihood and $P_{\text{old}}(\theta)$ is the prior. For $\beta = 0$ (no moment constraint) we recover Bayes' rule. For $\beta \neq 0$ Bayes' rule is modified by a "canonical" exponential factor.

## 6.2  Commuting and non-commuting constraints

The ME method allows one to process information in the form of constraints. When we are confronted with several constraints we must be particularly cautious. In what order should they be processed? Or should they be processed at the same time? The answer depends on the nature of the constraints and the question being asked.

We refer to constraints as *commuting* when it makes no difference whether they are handled simultaneously or sequentially. The most common example is that of Bayesian updating on the basis of data collected in multiple experiments: for the purpose of inferring $\theta$ it is well known that the order in which the observed data $x' = \{x'_1, x'_2, \ldots\}$ is processed does not matter. The proof that ME is completely compatible with Bayes' rule implies that data constraints implemented through $\delta$ functions, as in (6.2), commute just as they do in Bayes. It is useful to see how this comes about.

It is important to note that when an experiment is repeated it is common to refer to the value of $x$ in the first experiment and the value of $x$ in the second experiment. This is a dangerous practice because it obscures the fact that we are actually talking about *two* separate variables. We do not deal with a single $x$ but with a composite $x = (x_1, x_2)$ and the relevant space is $\mathcal{X}_1 \times \mathcal{X}_2 \times \Theta$. After the first experiment yields the value $x'_1$, represented by the constraint $c_1 : P(x_1) = \delta(x_1 - x'_1)$, we can perform a second experiment that yields $x'_2$ and is represented by a second constraint $c_2 : P(x_2) = \delta(x_2 - x'_2)$. These constraints $c_1$ and $c_2$ commute because they refer to *different* variables $x_1$ and $x_2$.[1]

In general constraints need not commute and when this is the case the order in which they are processed is critical. For example, suppose the prior is $P_{\text{old}}$ and we receive information in the form of a constraint, $C_1$. To update we maximize the entropy $S[P, P_{\text{old}}]$ subject to $C_1$ leading to the posterior $P_1$ as shown in Figure 6-1. Next we receive a second piece of information described by the constraint $C_2$. At this point we can proceed in essentially two different ways:

**(a) Sequential updating -** Having processed $C_1$, we use $P_1$ as the current prior and maximize $S[P, P_1]$ subject to the new constraint $C_2$. This leads us to the posterior $P_{\text{new}}^{(a)}$.

**(b) Simultaneous updating -** Use the original prior $P_{\text{old}}$ and maximize $S[P, P_{\text{old}}]$ subject to both constraints $C_1$ and $C_2$ simultaneously. This leads to the posterior $P_{\text{new}}^{(b)}$.[2]

---

[1] The use of the $\delta$ function has been criticized in that by implementing it, the probability is completely constrained, thus it cannot be updated by future information. This is certainly true! An experiment, once performed and its outcome observed, cannot be *un-performed* and its result cannot be *un-observed* by subsequent experiments. Thus, imposing constraint $c_1$ does not imply a revision of $c_2$.

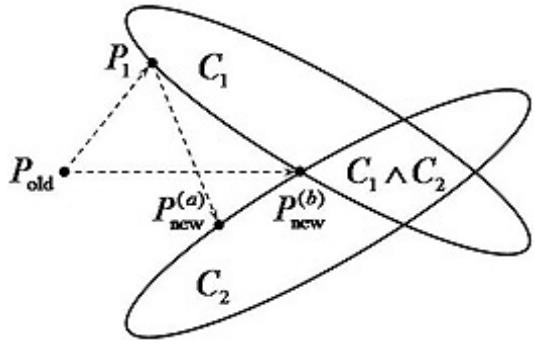

Figure 6-1: Illustrating the difference between processing two constraints $C_1$ and $C_2$ sequentially ($P_{\text{old}} \to P_1 \to P_{\text{new}}^{(a)}$) and simultaneously ($P_{\text{old}} \to P_{\text{new}}^{(b)}$ or $P_{\text{old}} \to P_1 \to P_{\text{new}}^{(b)}$).

To decide which path (a) or (b) is appropriate, we must be clear about how the ME method treats constraints. The ME machinery interprets a constraint such as $C_1$ in a very mechanical way: all distributions satisfying $C_1$ are in principle allowed and all distributions violating $C_1$ are ruled out.

---

[2] At first sight it might appear that there exists a third possibility of simultaneous updating: (c) use $P_1$ as the current prior and maximize $S[P, P_1]$ subject to both constraints $C_1$ and $C_2$ simultaneously. Fortunately, and this is a valuable check for the consistency of the ME method, it is easy to show that case (c) is equivalent to case (b). Whether we update from $P_{\text{old}}$ or from $P_1$ the selected posterior is $P_{\text{new}}^{(b)}$.

Updating to a posterior $P_1$ consists precisely in revising those aspects of the prior $P_\text{old}$ that disagree with the new constraint $C_1$. However, there is nothing final about the distribution $P_1$. It is just the best we can do in our current state of knowledge and we fully expect that future information may require us to revise it further. Indeed, when new information $C_2$ is received we must reconsider whether the original $C_1$ remains valid or not. Are *all* distributions satisfying the new $C_2$ really allowed, even those that violate $C_1$? If this is the case then the new $C_2$ takes over and we update from $P_1$ to $P_\text{new}^{(a)}$. The constraint $C_1$ may still retain some lingering effect on the posterior $P_\text{new}^{(a)}$ through $P_1$, but in general $C_1$ has now become obsolete.

Alternatively, we may decide that the old constraint $C_1$ retains its validity. The new $C_2$ is not meant to revise $C_1$ but to provide an additional refinement of the family of allowed posteriors. In this case the constraint that correctly reflects the new information is not $C_2$ but the more restrictive space where $C_1$ and $C_2$ overlap. The two constraints should be processed simultaneously to arrive at the correct posterior $P_\text{new}^{(b)}$.

To summarize: sequential updating is appropriate when old constraints become obsolete and are superseded by new information; simultaneous updating is appropriate when old constraints remain valid. The two cases refer to different states of information and therefore *we expect* that they will result in different inferences. These comments are meant to underscore the importance of understanding what information is being processed; failure to do so will lead to errors that do not reflect a shortcoming of the ME method but rather a misapplication of it.

## 6.3 Sequential updating: a loaded die example

This example, which considers a loaded die, illustrates the appropriateness of sequential updating. The background information is the following: A certain factory makes loaded dice. Unfortunately because of poor quality control, the dice are not identical and it is not known how each die is loaded. It is known, however, that the dice produced by this factory are such that face 2 is on the average twice as likely to come up as face 5.

The mathematical representation of this situation is as follows. The fact that we deal with dice is modeled in terms of multinomial distributions. The probability that casting a $k$-sided die $n$ times yields $m_i$ instances for the $i^{th}$ face is

$$P_{\text{old}}(m|\theta) = P_{\text{old}}(m_1...m_k|\theta_1...\theta_k, n) = \frac{n!}{m_1!...m_k!} \theta_1^{m_1}...\theta_k^{m_k} , \qquad (6.15)$$

where $m = (m_1, \ldots, m_k)$ with $\sum_{i=1}^{k} m_i = n$, and $\theta = (\theta_1, \ldots, \theta_k)$ with $\sum_{i=1}^{k} \theta_i = 1$. The generic problem is to infer the parameters $\theta$ on the basis of information about moments of $\theta$ and data $m'$. The additional information about how the dice are loaded is represented by the constraint $\langle \theta_2 \rangle = 2 \langle \theta_5 \rangle$. Note that this piece of information refers to the factory as a whole and not to any individual die. The constraint is of the general form of (6.1)

$$C_1 : \langle f(\theta) \rangle = F \quad \text{where} \quad f(\theta) = \sum_i^k f_i \theta_i . \qquad (6.16)$$

For this particular factory $F = 0$, and all $f_i = 0$ except for $f_2 = 1$ and $f_5 = -2$. Now that the background information has been given, here is our first example.

We purchase a die. On the basis of our general knowledge of dice we are led to write down a joint prior

$$P_{\text{old}}(m, \theta) = P_{\text{old}}(\theta) P_{\text{old}}(m|\theta) \ . \tag{6.17}$$

(The particular form of $P_{\text{old}}(\theta)$ is not important for our current purpose so for the sake of definiteness we can choose it flat.) At this point the only information we have is that we have a die and it came from a factory described by $C_1$. Accordingly, we use ME to update to a new joint distribution. This is shown as $P_1$ in Figure 6-1. The relevant entropy is

$$S[P, P_{\text{old}}] = -\sum_m \int d\theta \ P(x, \theta) \log \frac{P(x, \theta)}{P_{\text{old}}(x, \theta)} \ , \tag{6.18}$$

where

$$\sum_m = \sum_{m_1 \ldots m_k = 0}^n \delta(\textstyle\sum_{i=1}^k m_i - n) \quad \text{and} \quad \int d\theta = \int d\theta_1 \ldots d\theta_k \, \delta(\textstyle\sum_{i=1}^k \theta_i - 1) \ ,$$

Maximizing $S$ subject to normalization and $C_1$ gives the $P_1$ posterior

$$P_1(m, \theta) = \frac{e^{\lambda f(\theta)}}{Z_1} P_{\text{old}}(m, \theta) \ , \tag{6.19}$$

where the normalization constant $Z_1$ and the Lagrange multiplier $\lambda$ are determined from

$$Z_1 = \int d\theta \, e^{\lambda f(\theta)} P_{\text{old}}(\theta) \quad \text{and} \quad \frac{\partial \log Z_1}{\partial \lambda} = F \ . \tag{6.20}$$

The joint distribution $P_1(m, \theta) = P_1(\theta)P_1(m|\theta)$ can be rewritten as

$$P_1(m, \theta) = P_1(\theta)P_{\text{old}}(m|\theta) \quad \text{where} \quad P_1(\theta) = P_{\text{old}}(\theta)\frac{e^{\lambda f(\theta)}}{Z_1} . \tag{6.21}$$

To find out more about this particular die we toss it $n$ times and obtain data $m' = (m'_1, \ldots, m'_k)$ which we represent as a new constraint

$$C_2 : P(m) = \delta_{mm'} . \tag{6.22}$$

Our goal is to infer the $\theta$ that apply to our particular die. The original constraint $C_1$ applies to the whole factory while the new constraint $C_2$ refers to the actual die of interest and thus takes precedence over $C_1$. As $n \to \infty$ we expect $C_1$ to become less and less relevant. Therefore the two constraints should be processed sequentially.

Using ME, that is (5.10), we impose $C_2$ and update from $P_1(m, \theta)$ to a new joint distribution (shown as $P_{\text{new}}^{(a)}$ in Figure 6-1)

$$P_{\text{new}}^{(a)}(m, \theta) = \delta_{mm'}P_1(\theta|m) . \tag{6.23}$$

Marginalizing over $m$ and using (6.21) the final posterior for $\theta$ is

$$P_{\text{new}}^{(a)}(\theta) = P_1(\theta|m') = P_1(\theta)\frac{P_1(m'|\theta)}{P_1(m')} = \frac{1}{Z_2}e^{\lambda f(\theta)}P_{\text{old}}(\theta)P_{\text{old}}(m'|\theta) . \tag{6.24}$$

where

$$Z_2 = \int d\theta \, e^{\lambda f(\theta)} P_{\text{old}}(\theta) P_{\text{old}}(m'|\theta) . \tag{6.25}$$

The readers will undoubtedly recognize that (6.24) is precisely the result obtained by using MaxEnt to obtain a prior, in this case $P_1(\theta)$ given in (6.21), and then using Bayes' theorem to take the data into account. This familiar result has been derived in some detail for two reasons: first, to reassure the readers that ME does reproduce the standard solutions to standard problems and second, to establish a contrast with the example discussed next.

## 6.4 Simultaneous updating: a loaded die example

Here is a different problem illustrating the appropriateness of simultaneous updating. The background information is the same as in the previous example. The difference is that the factory now hires a quality control engineer who wants to learn as much as he can about the factory. His initial knowledge is described by the same prior $P_{\text{old}}(m, \theta)$, (6.17). After some inquiries he is told that the only available information is

$$C_1 : \langle \theta_2 \rangle = 2 \langle \theta_5 \rangle \ . \tag{6.26}$$

Not satisfied with this limited information he decides to collect data that reflect the production of the whole factory. He proceeds to collect $n$ randomly chosen dice, tossing each one once, yielding data $m' = (m'_1, \ldots, m'_k)$ which is represented as a constraint,

$$C_2 : P(m) = \delta_{mm'} \ . \tag{6.27}$$

The apparent resemblance with (6.22) may be misleading: (6.22) refers to a single die, while (6.27) now refers to $n$ dice. The goal here is to infer the distribution of $\theta$ that describes the overall population of dice produced by the factory. The new constraint $C_2$ is information in addition to, rather than instead of, the old $C_1$: the two constraints should be processed simultaneously. From (6.8) the joint posterior is[3]

$$P^{(b)}_{\text{new}}(m,\theta) = \delta_{mm'} P_{\text{old}}(\theta|m') \frac{e^{\beta f(\theta)}}{Z} \ . \tag{6.28}$$

Marginalizing over $m$ the posterior for $\theta$ is

$$P^{(b)}_{\text{new}}(\theta) = P_{\text{old}}(\theta|m') \frac{e^{\beta f(\theta)}}{Z} = \frac{1}{\zeta} e^{\beta f(\theta)} P_{\text{old}}(\theta) P_{\text{old}}(m'|\theta) \ , \tag{6.29}$$

where the new normalization constant is

$$\zeta = \int d\theta \, e^{\beta f(\theta)} P_{\text{old}}(\theta) P_{\text{old}}(m'|\theta) \quad \text{and} \quad \frac{\partial \log \zeta}{\partial \beta} = F \ . \tag{6.30}$$

This looks similar to the sequential case, (6.24), but there is a crucial difference: $\beta \neq \lambda$ and $\zeta \neq Z_2$, where $\lambda$ is determined by (6.20), $\beta$ is determined by (6.29). In the sequential updating case, the multiplier $\lambda$ is chosen so that the intermediate $P_1$ satisfies $C_1$ while the posterior $P^{(a)}_{\text{new}}$ only satisfies $C_2$. In the simultaneous updating case the multiplier $\beta$ is chosen so that the posterior $P^{(b)}_{\text{new}}$ satisfies both $C_1$ and $C_2$ or $C_1 \wedge C_2$. Ultimately, the two distributions $P_{\text{new}}(\theta)$ are different because they refer to different problems: $P^{(a)}_{\text{new}}(\theta)$ refers to a single die, while $P^{(b)}_{\text{new}}(\theta)$ applies to the set of $n$ dice obtained from the factory.[4]

---

[3] As mentioned in the previous footnote, whether we update from $P_{\text{old}}$ or from $P_1$ we obtain the same posterior $P^{(b)}_{\text{new}}$.

[4] For the sake of completeness, we note that, because of the peculiarities of $\delta$ functions, had the constraints been processed sequentially but in the opposite order, first the data $C_2$, and then the moment $C_1$, the resulting posterior would be the same as for simultaneous update to $P^{(b)}_{\text{new}}$.

## 6.5 More on the multinomial problem

In this section, we show that when the data is discrete, we can produce a compact form for the entropy that is similar to the thermodynamic entropy. We also show how the posterior of the multinomial problem (6.29) can be calculated.

**Compact form of the multinomial entropy**

Here we calculate another form for the above entropy. We start with start with the general posterior,

$$P_{\text{new}}(m, \theta) = P_{\text{old}}(m, \theta)\delta_{mm'}\frac{e^{\beta f(\theta)}}{\zeta(x, \beta)} \ . \tag{6.31}$$

where $\zeta(m, \beta) = \int d\theta e^{\beta f(\theta)} P_{\text{old}}(m, \theta)$. This is substituted into the appropriate entropy (6.18) which yields,

$$S = -\sum_m \int d\theta \ P_{\text{old}}(m, \theta)\delta_{mm'}\frac{e^{\beta f(\theta)}}{\zeta(m, \beta)} \log \frac{P_{\text{old}}(m, \theta)\delta_{mm'}\frac{e^{\beta f(\theta)}}{\zeta(m,\beta)}}{P_{\text{old}}(m, \theta)} \ . \tag{6.32}$$

We cancel the priors in the log,

$$S = -\sum_m \int d\theta \ P_{\text{old}}(m, \theta)\delta_{mm'}\frac{e^{\beta f(\theta)}}{\zeta(m, \beta)} \log \frac{\delta_{mm'}e^{\beta f(\theta)}}{\zeta(m, \beta)} \ , \tag{6.33}$$

then sum over the discrete $m$ which yields,

$$S = -\int d\theta \ P_{\text{old}}(m', \theta)\frac{e^{\beta f(\theta)}}{\zeta'(m', \beta)} \log \frac{e^{\beta f(\theta)}}{\zeta'(m', \beta)} \ . \tag{6.34}$$

Notice that $\zeta(x, \beta)$ is a function of $x$. Thus when the delta function is evaluated, $\zeta(m, \beta) \to \zeta(m', \beta)$. This can be rewritten as

$$S = -\int d\theta \; P_{\text{old}}(m', \theta) \frac{e^{\beta f(\theta)}}{\zeta'(m', \beta)} \left( \log e^{\beta f(\theta)} - \log \zeta'(m', \beta) \right) . \tag{6.35}$$

Since $\beta$ and $\log \zeta'(m', \beta)$ are not functions of $\theta$, this can then be written as

$$S = -\frac{\beta}{\zeta'(m', \beta)} \int d\theta \; P_{\text{old}}(m', \theta) e^{\beta f(\theta)} f(\theta) + \log \zeta'(m', \beta) \int d\theta \; P_{\text{old}}(m', \theta) \frac{e^{\beta f(\theta)}}{\zeta'(m', \beta)} . \tag{6.36}$$

Notice, since $\zeta(x', \beta) = \int d\theta e^{\beta f(\theta)} P_{\text{old}}(x', \theta)$, the second term becomes, $\log \zeta'$. Also, since $\int d\theta \; P_{\text{old}}(x', \theta) e^{\beta f(\theta)} f(\theta) = \partial \zeta'/\partial \beta = F \zeta'$, the first term is $-\beta F$ and thus the entropy can now be written as

$$S[P, P_{\text{old}}] = -\beta F + \log \zeta'(x', \beta) . \tag{6.37}$$

Shannon had discovered the entropy that bears his name quite independently of thermodynamic considerations. Jaynes extended Shannon's logic by using a generalization of his method and derived the canonical ensemble of Gibbs that is used in thermodynamics. In our method we generalize both Shannon and Jaynes' logic (which has nothing to do with thermodynamics directly) yet our new measure for diversity turns out to take the same form as the thermodynamic entropy. The realization that the ME diversity is of the exact same form as the thermodynamic may be useful. Concepts in thermodynamics such as energy, could be used applied to other fields.[5]

---

[5]The thermodynamic entropy is actually, $S = \log \zeta + \beta F$. The fact that our entropy (6.37) has a $-\beta F$ is a reflection of our choice to add our Lagrange multipliers in (6.3) as opposed to subtracting

**Calculation of the multinomial posterior**

Here we pursue the calculation of the posterior (6.29) in more detail. To be specific we choose a flat prior, $P_{\text{old}}(\theta) = \text{constant}$. Then, dropping the superscript (b),

$$P_{\text{new}}(\theta) = \frac{1}{\zeta_e} \delta(\sum_i \theta_i - 1) \prod_{i=1}^{k} e^{\beta f_i \theta_i} \theta_i^{m'_i}. \tag{6.38}$$

where $\zeta$ is,

$$\zeta = \int \delta(\sum_i \theta_i - 1) \prod_{i=1}^{k} d\theta_i e^{\beta f_i \theta_i} \theta_i^{m'_i}, \tag{6.39}$$

and $\beta$ is determined from (6.30) which in terms of $\zeta$ now reads $\partial \log \zeta / \partial \beta = F$. A brute force calculation gives $\zeta$ as a nested hypergeometric series,

$$\zeta = e^{\beta f_k} I_1(I_2(\ldots(I_{k-1}))), \tag{6.40}$$

where each $I$ is written as a sum,

$$I_j = \Gamma(b_j - a_j) \sum_{q_j=0}^{\infty} \frac{\Gamma(a_j + q_j)}{\Gamma(b_j + q_j) \, q_j!} t_j^{q_j} I_{j+1} \quad \text{with} \quad I_k = 1 \,. \tag{6.41}$$

The index $j$ takes all values from 1 to $k-1$ and the other symbols are defined as follows: $t_j = \beta(f_{k-j} - f_k)$, $a_j = m'_{k-j} + 1$, and

$$b_j = n + j + 1 + \sum_{i=0}^{j-1} q_i - \sum_{i=0}^{k-j-1} m'_i \,, \tag{6.42}$$

---

them as Jaynes did to get the thermodynamic entropy. However, this is trivial because when one solves for $\beta$ in (6.12) the sign will be accounted for. In other words, if the Lagrange multiplier was subtracted, the solution to (6.12) would be $-F$ and the entropy would have a $+\beta F$.

with $q_0 = m'_0 = 0$. The terms that have indices $\leq 0$ are equal to zero (i.e. $b_0 = q_0 = 0$, etc.). A few technical details are worth mentioning: First, one can have singular points when $t_j = 0$. In these cases the sum must be evaluated as the limit as $t_j \to 0$. Second, since $a_j$ and $b_j$ are positive integers the gamma functions involve no singularities. Lastly, the sums converge because $a_j > b_j$. The normalization for the first die example, (6.25), can be calculated in a similar way. Currently, for small values of $k$ (less than 10) it is feasible to evaluate the nested sums numerically; for larger values of $k$ it is best to evaluate the integral for $\zeta$ using sampling methods.

## 6.6 Summary

The realization that the ME method incorporates Bayes' rule as a special case has allowed us to go beyond Bayes' rule to process both data and expected value constraints simultaneously. To put it bluntly, anything one can do with Bayes can also be done with ME with the additional ability to include information that was inaccessible to Bayes alone. This raises several questions and we have offered a few answers.

First, it is not uncommon to claim that the non-commutability of constraints represents a *problem* for the ME method. Processing constraints in different orders might lead to different inferences and this is said to be unacceptable. We have argued that, on the contrary, the information conveyed by a particular sequence of constraints is not the same information conveyed by the same constraints in different order. Since different informational states should in general lead to different inferences, the way ME handles non-commuting constraints should not be regarded as a *shortcoming* but rather as a *feature* of the method.

Why is this type of problem beyond the reach of Bayesian methods? After all, we can always interpret an expected value as a sample average in a sufficiently large number of trials. This is equivalent to constructing a large imaginary ensemble of experiments. Entropy methods then become in principle *superfluous*; all we need is probability. The problem with inventing *imaginary* ensembles to do away with entropy in favor of mere probabilities, or to do away with probabilities in favor of more intuitive frequencies, is that the ensembles are just what they are claimed to be, imaginary. They are purely artificial constructions invented for the purpose of handling incomplete information. It seems to us that a safer way to proceed is to handle the available information directly as given (i.e., as expected values) without making additional assumptions about an imagined reality.

Finally we address the title of this chapter: Universal updating. The data, constraints, priors and posteriors are of a general form in the first section. However, the *process* that was used to produce our result is *universal*. It consistently encompasses all methods of inference.

# Chapter 7

# Potential applications using ME

## 7.1 Applying ME in ecology

The core of this chapter has been previously published in [46].

Diversity is a concept that is used in many fields to describe the variability of different entities in a group. In ecology, the Shannon index [47] and Simpson's index [48] are the predominant measures of diversity. The purpose of measuring diversity is to judge the relationship between other communities or to communities and their environmental conditions. In this paper we focus on the Shannon index since it is closely tied to many other areas of research, such as information theory and physics.

It is often the case that the individuals comprising a community cannot be fully counted. In this case, when one has incomplete information, one must rely on methods of inference. The purpose of this section is to demonstrate how the ME method can be used in the measure of diversity. By using ME, we have the ability to include more information than the traditional use of Shannon's measure allows.

Traditionally when confronted with a community whose individuals cannot be fully counted, the frequencies of the species that are counted are used to calculate the diversity. The frequency is used because it represents an estimate of the probability of finding a particular species in the community. However, the frequency is not equivalent to the probability (shown in chapter 1) and may be a poor estimate. Fortunately, there are much better methods for estimating or inferring the probability such as MaxEnt and Bayes. Even more fortunate is that the new ME method can reproduce every aspect of Bayesian and MaxEnt inference *and* tackle problems that the two methods alone could not address.

We examine one of the sets of axioms describing the desirable attributes of a diversity measure [47]. The attributes will single out the Shannon entropy as the proper measure. We then solve a toy ecological problem and discuss the diversity calculated by using the traditional approach and the diversity calculated by incorporating the ME method. This illustrates the many advantages to using the ME method with out losing the desirable properties of the traditional diversity measure.

**Traditional Diversity**

In 1969, Evelyn Christine Pielou described the properties that a measure of diversity, $H$ should possess [47]:

### Pielou's Axioms

1. For a given number of species, $s$, $H$ should have its greatest value when $p_i = 1/s$ for all $i$. Such a community will be called *completely even*.

2. Given two completely even communities, one with $s$ species and the other with

$s+1$, the latter should have the *greater* value of $H$.

3. If individuals in a community are classifiable in more than one way, then one should be able to separate out the classifications from the joint classification.

For the 3<sup>rd</sup> axiom, we quote Pielou, "Suppose the community members are subject to two separate classifications (not necessarily independent), namely an $A$-classification with $a$ classes and a $B$-classification with $b$ classes. Let $p_i (i = 1, \ldots, a)$ be the proportion of community members in the $i^{th}$ class of the $A$-classification; let $q_{ij}(i = 1, \ldots, a;\ j = 1, \ldots, b)$ be the proportion of these members that belong to the $j^{th}$ class of the $B$-classification. And put $p_i q_{ij} = \pi_{ij}$ so that $\pi_{ij}$ is the proportion of the whole community that belongs to the $i^{th}$ $A$-class and $j^{th}$ $B$-class.

Also, put $H(AB)$ for the diversity of the doubly classified community; $H(A)$ for the diversity under the $A$-classification only; and $H(B)$ for the diversity under the $B$-classification of that part of the community belonging to the $i^{th}$ $A$-class.

Let $H_A(B) = \sum p_i H_i(B)$ be the mean of the $H_i(B)$ over all $A$-classes.

We then require that $H(AB) = H(A) + H_A(B)$."

Pielou showed that the only measure that satisfied these axioms was

$$H = -C \sum_i^k p_i \log p ,\qquad(7.1)$$

where $C$ is a positive constant, $p_i = m_i/n$, $m_i$ represents the counts of each species and $n$ represents the total of all species so that $n = \sum_i^k m_i$. This was called the Shannon index or Shannon diversity because Claude Shannon used a similar set of axioms to attain the identical function (as shown in Chapter 3).

**Toy Example**

The general information for our example is as follows: There are $k$ known types of plants in a forest. A portion of the forest is examined and the number of each species is counted where $m_1, m_2 \ldots m_k$ represents the counts of each species and $n$ represents the total count so that $n = \sum_i^k m_i$. Additionally, we know from biological examination that one species, $s_2$ and another species, $s_5$ are codependent. Perhaps they each need something from the other in order to function. For simplicity, let this dependence be such that on the average, twice the number of $s_2$ will be found compared to $s_5$.

The problem with a method using the function (7.1) as a measure of diversity is not in the method itself but with the reason it is being used. If the purpose of using this method is to measure the diversity of the portion of the community that was counted or to measure the diversity of a community that is fully censused (fully counted), then the method is acceptable. However, if the purpose of the method is to estimate or infer the diversity of a community that *cannot* be fully counted (i.e. the whole forest), then it may be a poor estimate. First, $p_i$ is meant to represent the probability of finding the $i^{th}$ species in the forest. As shown in chapter 1, the frequency of the sample, $\nu$ is not equivalent to the probability, $p$. In fact, it is the expected value of the frequency that is equivalent to the probability, $\langle \nu \rangle = p$. It would only make sense to use the frequency as an estimate of the probability when the number of samples or counts, $n$ is very large (i.e. $n \to \infty$) but this is not usually the case. Second, there is no clear way to process the information about the codependence using Shannon's entropy. We would wish to include more information than just the frequency, such as other constraints like codependence.

**Applying ME to diversity**

Here we intend to use a better method to estimate or infer $p_i$ and that method is the ME method. Since this probability is unknown, we will use $\theta_i$ as a parameter that represents the probability. The first task is to realize that the correct mathematical model for the probability of getting a particular species where the information that we have is the number of species counted. The proper model is the multinomial distribution. For our example, the probability of finding $m_i$ instances of the $i^{th}$ species, where the total number of species, $k$ and the total number of counts, $n$ are known is

$$P_{\text{old}}(m|\theta, n) = \frac{n!}{m_1! \ldots m_k!} \theta_1^{m_1} \ldots \theta_k^{m_k} , \tag{7.2}$$

where $m = (m_1, \ldots, m_k)$ with $\sum_{i=1}^{k} m_i = n$, and $\theta = (\theta_1, \ldots, \theta_k)$ with $\sum_{i=1}^{k} \theta_i = 1$. The general problem is to infer the parameters $\theta$ on the basis of information about the data, $m$ and other possible constraints, such as codependency. Here we see the first advantage with using the ME diversity; we allow for fluctuations in our inference by looking at a distribution of $\theta$s as opposed to claiming that we know the "true" $\theta$.

Additionally we can include information about the codependence by using the following general constraint,

$$\langle f(\theta) \rangle = F \quad \text{where} \quad f(\theta) = \sum_{i}^{k} f_i \theta_i , \tag{7.3}$$

where $f_i$ is used to represent the codependence. For our example, on the average, we will find twice the number of $s_2$ as compared to $s_5$ thus, *on the average*, the probability of finding one of the species will be twice that of the other, $\langle \theta_2 \rangle = 2 \langle \theta_5 \rangle$. In this case,

$f_2 = 1$, $f_5 = -2$ and $f_{i \neq (2,5)} = F = 0$.

Next we need to write the observed data (counts) as a constraint which in general is

$$P(m|n) = \delta_{mm'} \, , \tag{7.4}$$

where $m' = \{m'_1, \ldots, m'_k\}$. Finally we write the appropriate entropy,

$$S[P, P_{\text{old}}] = -\sum_m \int d\theta P(m, \theta|n) \log \frac{P(m, \theta|n)}{P_{\text{old}}(m, \theta|n)} \, , \tag{7.5}$$

where

$$\sum_m = \sum_{m_1 \ldots m_k = 0}^{n} \delta(\sum_{i=1}^{k} m_i - n) \, , \tag{7.6}$$

and

$$\int d\theta = \int d\theta_1 \ldots d\theta_k \, \delta \left( \sum_{i=1}^{k} \theta_i - 1 \right) \, , \tag{7.7}$$

and where $P_{\text{old}}(m, \theta|n) = P_{\text{old}}(\theta|n) P_{\text{old}}(m|\theta, n)$. We then maximize this entropy with respect to $P(m, \theta|n)$ subject to normalization and our constraints which after marginalizing over $m'$ yields,

$$P(\theta|n) = P_{\text{old}}(\theta|n) P_{\text{old}}(m'|\theta, n) \frac{e^{\beta f(\theta)}}{\zeta} \, , \tag{7.8}$$

where

$$\zeta = \int d\theta \, e^{\beta f(\theta)} P_{\text{old}}(\theta|n) P_{\text{old}}(m'|\theta, n) \quad \text{and} \quad F = \frac{\partial \log \zeta}{\partial \beta} \, . \tag{7.9}$$

The probability distribution $P(\theta|n)$ has sometimes been criticized for being too strange. The idea of getting a 'probability of a probability' may seem strange at first but makes absolute sense. We do not know the "true" distribution of species, $\theta_i$. Therefore it seems natural to express our knowledge with some uncertainty in the form of a distribution. Notice that if one has no information relating the species then $\beta = 0$.

Now that we have a distribution of possible probabilities for each species, which one should we choose? Often the mean of the distribution is used chosen as a representative estimate. We then use this estimate in the diversity measure so that we have,

$$H_{ME} = -\sum_i^k \langle \theta_i \rangle \log \langle \theta_i \rangle \ , \qquad (7.10)$$

where $C = 1$ and $\langle \theta_i \rangle = \int d\theta \ \theta_i \ P(\theta|n)$ is the mean of (7.8). Another possible solution would be the mean of the diversity measure itself

$$\langle H \rangle = \int d\theta P(\theta|n) \left( -\sum_i^k \theta_i \log \theta_i \right) \ . \qquad (7.11)$$

**Summary**

Diversity is an important concept in many fields, such as ecology. In this paper we provided a toy example of how ME would be used to produce a better estimate that could be used in a measure of diversity, as opposed to using a frequency. By using the multinomial, we not only properly infer $\theta$ so that fluctuations are represented, we get the bonus of being able to include additional constraint information.

## 7.2 Complex Agents

The core of this chapter has been previously published in [49].

There are many examples of systems where agents respond to both local information as well as global information. Nature yields many such examples where cells react to local stimuli yet carry some global instructions, such as reproduction. The examples get more complex when the cells interact locally or share information. We would like to infer something about the system or better, what each *agent* infers about the system. It is this latter case that we will be specifically addressing. The main purpose of this section is to examine a situation where each agent in a network (of varying degrees of complexity) infers something about the whole system based on limited information.

We solve a toy problem where we include global information in the form of a moment constraint or expected value and then introduce local information in the form of data. This will show how the agents infer aspects of the whole system using the same process yet come to different conclusions. Complexity is increased as the number of agents are increased yet the complexity of the process does not grow proportionately. This illustrates the advantages to using the ME method.

**The agent example**

Let us start with a very simple example: There is a class with 3 students sitting in desks next to each other and one professor. The professor announces that he has a machine that produces loaded, 3 sided dice and he would like his students to try to infer the probability of each getting a 1, a 2 or a 3 in a single toss. He tells them that he has created this machine in such a way that *on the average*, a side 1 is twice as likely to come up as a side 3. Now he rolls the a die without showing them the results. He announces that he has rolled $n$ of these dice, each once. Then he writes down how many times a 1 came up on a piece of paper and hands it to student A, careful not to let the other students see it. He proceeds to do this for each of the other students, giving student B the results of side 2 and student C the results of side 3. What would each student determine the probabilities of the sides of the next die to be created be? Each needs to determine the probability of getting *any* particular outcome of a die that is produced in one toss ($\theta_i$) given the information.

We summarize the information the following way: there are 3 agents, A, B and C. Randomly chosen dice are rolled once with the counts of each side that is rolled represented by, $m_1, m_2$ and $m_3$ respectively with $n$ representing the total number of rolls (as well as dice since each dice is only tossed once) so that $n = \sum_{i=1}^{3} m_i$. Additionally, we know that on the average one a side, $\theta_1$ is twice as likely to be rolled as $\theta_3$, or better, $\langle \theta_1 \rangle = 2 \langle \theta_3 \rangle$

The first task is to realize that the correct mathematical model for the probability of getting a particular side where the information that we have is the number of sides counted is a multinomial distribution. The probability of finding $k$ sides in $n$ counts which yields $m_i$ instances for the $i^{th}$ side is

$$P_{\text{old}}(m|\theta, n) = \frac{n!}{m_1! \ldots m_k!} \theta_1^{m_1} \ldots \theta_k^{m_k}, \qquad (7.12)$$

where $m = (m_1, \ldots, m_k)$ with $\sum_{i=1}^{k} m_i = n$, and $\theta = (\theta_1, \ldots, \theta_k)$ with $\sum_{i=1}^{k} \theta_i = 1$. The general problem is to infer the parameters $\theta$ on the basis of information about the data, $m$.

Additionally we can include information about the bias of the sides by using the following general constraint,

$$\langle f(\theta) \rangle = F \quad \text{where} \quad f(\theta) = \sum_i^k f_i \theta_i \;, \tag{7.13}$$

where $f_i$ is used to represent the machine bias. For our example, on the average, we will find twice the number of $\theta_1$ as compared to $\theta_3$ thus, *on the average*, the probability of finding one of the sides will be twice that of the other, $\langle \theta_1 \rangle = 2 \langle \theta_3 \rangle$. In this case, $f_1 = 1$, $f_3 = -2$ and $f_2 = F = 0$.

Next we need to write the observed data (counts) as a constraint which for student A is

$$P(m_1|n) = \delta_{m_1 m_1'} \;, \tag{7.14}$$

where $m_1'$ is the number of counts for side 1. The appropriate form of the entropy is

$$S[P, P_{\text{old}}] = -\sum_m \int d\theta P(m, \theta|n) \log \frac{P(m, \theta|n)}{P_{\text{old}}(m, \theta|n)} \;, \tag{7.15}$$

where

$$\sum_m = \sum_{m_1 \ldots m_3 = 0}^n \delta(\sum_{i=1}^3 m_i - n) \;, \tag{7.16}$$

and

$$\int d\theta = \int d\theta_1 d\theta_2 \delta\left(\sum_{i=1}^3 \theta_i - 1\right) \;, \tag{7.17}$$

and where $P_{\text{old}}(m, \theta|n) = P_{\text{old}}(\theta|n) P_{\text{old}}(m|\theta, n)$. The prior $P_{\text{old}}(\theta)$ is not important for our current purpose so for the sake of definiteness we can choose it flat for our example (there are most likely better choices for priors). We then maximize this entropy (7.15) with respect to $P(m, \theta|n)$ subject to normalization and our constraints to get our joint posterior,

$$P_{\text{A}}(\theta, m) = \frac{\delta_{m_1 m_1'}}{\zeta_{\text{A}}} \frac{n!}{m_1! m_2! (n-m_1-m_2)!} \theta_1^{m_1} \theta_2^{m_2} (1-\theta_1-\theta_2)^{n-m_1-m_2} e^{\beta(3\theta_1+2\theta_2-2)} \quad (7.18)$$

where $\zeta_{\text{A}}$ is a normalization constant. After marginalizing over $m$ we get,

$$P_{\text{A}}(\theta_1, \theta_2) = \frac{1}{\zeta_{\text{A}}'} e^{\beta(3\theta_1+2\theta_2-2)} \theta_1^{m_1'} (1-\theta_1)^{n-m_1'} , \quad (7.19)$$

where

$$\zeta_{\text{A}}' = \int d\theta_1 d\theta_2 \, e^{\beta(3\theta_1+2\theta_2-2)} \theta_1^{m_1'} (1-\theta_1)^{n-m_1'} , \quad (7.20)$$

and $\beta$ is determined from

$$F = \frac{\partial \log \zeta}{\partial \beta} .$$

Notice that if one has no information relating the sides then $\beta = 0$.

This is the probability distribution that student A would assign to the die. Since all of the students will follow the same proper inference method (ME), we need only look at one of the student's solutions in general. The only difference between each students' solution will be the value of $m'$. Notice that all students or agents agree on some global information, the bias of the die machine and the number of dice rolled. However, in general they will determine a different probability distribution that is dependent on the local information, $m'$, in this case the total number of a particular side.

Let us now complicate the problem further. Imagine that each student's desk is at a vertex of an equilateral triangle (so that they are equidistant from each other).

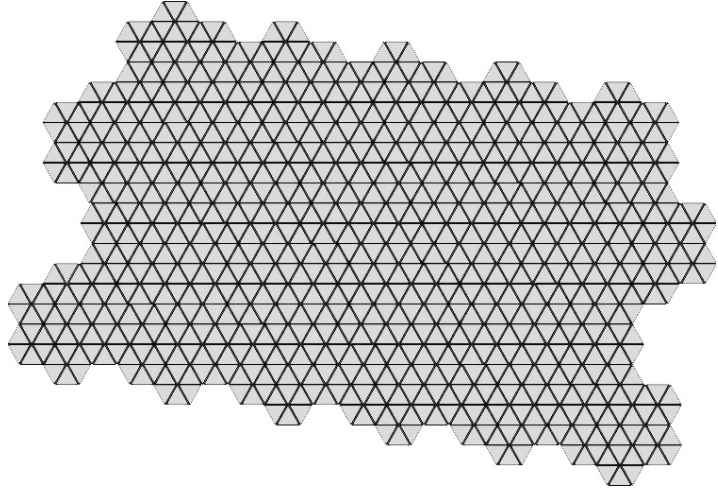

Figure 7-1: An example structure that relates agents in a system. Here each vertex is an agent.

They notice that the teacher is looking the other way so they each glance at their neighbor's paper. Since each of them now have *all* of the information they should all come up with the same answers. However, let us increase increasing the number of students. We enlarge the class by adding $k$ students with a professor rolling a $k$ sided dice that are loaded in some given way. The students are arranged in a lattice structure such as in Figure 7-1. where there is one student at each of the vertices. Each student that is not on an boundary now has six neighbors. Thus if they are allowed to 'look' at their nearest neighbors, the form of the probability distribution that each student would now assign is

$$P_{\rm S}(\theta_1...\theta_{k-1}) = \frac{1}{\zeta_{\rm S}} e^{\beta f_k \left(1 - \sum\limits_{i}^{k-1} \theta_i\right)} (1 - \sum_i \theta_i)^{n - \sum\limits_{i}^{7} m'_i} \prod_{i=1}^{7} \theta_i^{m'_i} \prod_{i=1}^{k-1} e^{\beta f_i \theta_i} \ . \qquad (7.21)$$

given the constraints (7.13) and $P(m_1|n) = \delta_{m_1 m'_1}, \ldots, P(m_7|n) = \delta_{m_7 m'_7}$.

**Summary**

We demonstrated that the ME method can easily lend itself to agent-based modeling. Whether the agents are skin cells, banks in a network or students in a classroom, the methodology of ME can be applied in order to model many of these systems. Any system where agents agree on some global information yet react to local information should be able to be modeled with this method. The observed local data is applied as a constraint such as (7.13) within a model such as (7.12) and the global information is applied in a constraint such as (7.14). This particular toy problem is complex in the sense that there can be a very large number of agents. By determining what each agent 'sees' we can predict many properties of the system. A future application in this regard would be to apply decision theory concepts to this approach.

# Chapter 8

# Final remarks and future applications

In part 1, we detailed how we arrived at our present understanding of probabilities and how we manipulate them – the product and addition rules by Cox. We also discussed the modern view of entropy and how it relates to known entropies such as the thermodynamic entropy and the information entropy. We noted that entropy, including that of Gibbs, is not dependent on an equilibrium situation. This led to the method of maximum entropy, MaxEnt. It was pointed out throughout that the results of the methods depended on two key points: First, the methods are a result of *consistency*. Second, the methods reflect the information that is put into them.

In chapter 4, we showed that Skilling's method of induction led us to a unique general theory of inductive inference, the ME method. The whole approach is extremely conservative. First, the axioms merely instruct us what not to update – do not change your mind except when forced by new information. Finally, we went beyond the insights of Karlbelkar and Uffink, and showed that our consistency axiom selects a unique, universal value for the parameter $\eta$ and this value ($\eta = 0$) corresponds to the usual logarithmic entropy. The advantage of our approach is that it shows precisely how it is that $\eta$-entropies with $\eta \neq 0$ are ruled out as tools for updating, such as those of Renyi or Tsallis.

In chapter 5, we explored the compatibility of Bayes and ME updating. After pointing out the distinction between Bayes' theorem and the Bayes' updating rule, we showed that Bayes' rule is a special case of ME updating by translating information in the form of data into constraints that can be processed using ME. This implies that ME is *capable of reproducing every aspect of orthodox Bayesian inference* and proves the complete compatibility of Bayesian and entropy methods. We illustrated this by showing that ME can be used to derive two results traditionally in the domain of Bayesian statistics, Laplace's Succession rule and Jeffrey's conditioning rule.

In chapter 6, we brought the other chapters together to show that ME is the *universal* method for inference. The realization that the ME method incorporates Bayes' rule as a special case has allowed us to go beyond Bayes' rule and to process both data and expected value constraints simultaneously. To put it bluntly, anything one can do with Bayes can also be done with ME with the additional ability to include information that was inaccessible to Bayes alone. The generic "canonical" form of the posterior distribution for the problem of simultaneous updating with data and moments was obtained.

We discussed the general problem of non-commuting constraints, when they should be processed sequentially and when simultaneously. First, it is not uncommon to hear the claim that the non-commutability of constraints represents a *problem* for the ME method. Processing constraints in different orders might lead to different inferences and this is said to be unacceptable. We have argued that, on the contrary, the information conveyed by a particular sequence of constraints is not the same information conveyed by the same constraints in different order. Since different informational states should in general lead to different inferences, the way ME handles non-commuting constraints should not be regarded as a *shortcoming* but rather as a *feature* of the method. To illustrate this, a multinomial example of die tosses is solved in detail for two superficially similar but actually very different problems.

In chapter 7, we attempt to show that ME is truly universal by applying it to other problems outside of physics, from which it originates. The applications reflect problems of potential interest. In ecology, we reexamine how diversity is measured and improve upon the current method by allowing for fluctuations and not relying on asymptotic arguments. In the agent application, we discuss how the ME method can be used to model complex systems that have local information as well as global information.

# Bibliography

[1] J. Bernoulli, *The Art of Conjecturing, together with Letter to a Friend on Sets in Court Tennis*, English trans. E. D. Sylla (Johns Hopkins, 2005); *Ars Conjectandi: Usum & Applicationem Praecedentis Doctrinae in Civilibus*, (Moralibus & Oeconomicis, 1713).

[2] A. Arnauld, *Logic, or the Art of Thinking*, English trans., J. Vance Buroker, (Cambridge 1996); *La Logique ou l'Art de Penser*, (Paris 1683).

[3] S. D. Poisson, *Treatise on Mechanics,* (London in 1842); "Recherches sur la probabilité des jugements, principalement enmatière criminelle", *C.r. Acad. Sci. Paris*, **t1,** 473, (1835); "Note sur la loi des grands nombres", Ibidem, **t2**, 377, (1836).

[4] J. Uffink, Stud. Hist. Phil. Mod. Phys. **26B**, 223 (1995).

[5] T. Bayes, "An Essay towards solving a Problem in the Doctrine of Chances", *Phil. Trans. Roy. Soc*, 330, (1763).

[6] P. S. Laplace, English edition, *A Philosophical Essay on Probabilities*, (New York, Dover, 1951); *Theorie Analytique des Probabilities*, 2 vols.(1812).

[7] E. T. Jaynes, "Bayesian Methods: General Background", *Maximum Entropy and Bayesian Methods in Applied Statistics*, ed. J. H. Justice, pp1-25, (Cambridge, 1985).

[8] R. T. Cox , "Probability, Frequency, and Reasonable Expectation", *Am. J. Phys.* **17**, 1, (1946); *The Algebra of Probable Inference*, (Johns Hopkins University Press 1961).

[9] A. N. Kolmogorov, *Foundations of Probability*, (Chelsea, 1950); *Grundbegriffe der Wahrscheinlichkeitrechnung, Ergebnisse Der Mathematik*, (1933).

[10] K. Knuth, "Deriving Laws from Ordering Relations", *Bayesian Inference and Maximum Entropy Methods in Science and Engineering*, eds. G. J. Erickson and Y. Zhai, AIP Conf. Proc. **707**, 75 (2004).

[11] J. Uffink, "Boltzmann's Work in Statistical Physics", *Stanford Encyclopedia of Philosophy*, (2004).

[12] J. C. Maxwell, "Illustrations of the Dynamical Theory of Gases.", (1859); *Collected Works*, ed. W. D. Niven, **I** ,377, (London 1890).

[13] A. Caticha, Class notes on "Principles of Information Physics", (Univ at Albany - SUNY, 2007).

[14] R. J. E. Clausius, "On the Moving Force of Heat, and the Laws regarding the Nature of Heat itself which are deducible therefrom", English trans, *Phil. Mag.* **2**, 1–21, 102–119, (1851); "Über die bewegende Kraft der Wärme, Part I, Part II", *Annalen der Physik* **79**: 368–397, 500–524, (1850).

[15] J. Uffink, "Boltzmann's Work in Statistical Physics", *Stanford Encyclopedia of Philosophy*, (2004) references L. Boltzmann, *Wiener Berichte* **76**, 373, (1877).

[16] R. C. Tolman, *The Principles of Statistical Mechanics* (Oxford University Press, London, 1938).

[17] M, Planck, "On the Law of Distribution of Energy in the Normal Spectrum", *Annalen der Physik*, **4**,553, (1901).

[18] J. W. Gibbs, *Elementary Principles in Statistical Mechanics*, (1902); Reprinted in Collected works and commentary, (Yale University Press 1936), and (Dover Publications, Inc. 1960).

[19] E. T. Jaynes, "Gibbs vs Boltzmann Entropies," *Am. J. Phys.* **33**, 391, (1965).

[20] E. T. Jaynes, "The Evolution of Carnot's Principle", *Bayesian Inference and Maximum Entropy Methods in Science and Engineering*, eds. G. J. Erickson and C. R. Smith, pp 267-281, (Kluwer 1988).

[21] C. E. Shannon, "A Mathematical Theory of Communication", *Bell System Technical Journal*, **27**, 379, (1948).

[22] E. T. Jaynes, Phys. Rev. **106**, 620 and **108**, 171 (1957).

[23] T. M. Cover and J. A. Thomas, *Elements of Information Theory - 2nd Ed.* (Wiley, New York 2006).

[24] L. Brillouin, *Science and Information Theory*, (Academic Press, New York, 1956).

[25] H. Jeffreys, *Theory of Probability*, (Oxford, 1939), (Later editions, 1948, 1961, 1979).

[26] S. Kullback. *Information Theory and Statistics*. (Wiley, 1959)

[27] E. T. Jaynes, "Information Theory and Statistical Mechanics," *Statistical Physics*, ed. K.W. Ford, W. A. Benjamin, Inc., 181.(1968).

[28] F. P. Ramsey, *Truth and Probability In The Foundations of Mathematics and Other Logical Essays*, ed R. B. Braithwaite, Routledge and Kegan Paul, London, 156-198, (1931).

[29] L. J. Savage, *Foundations of Statistics*, (Dover, New York, 1954).

[30] B. De Finetti, *Probability, Induction, and Statistics: The Art of Guessing*, (Wiley, New York, 1972).

[31] A. Wald, *Statistical Decision Functions*, (Wiley, New York, 1950).

[32] K. Knuth, "Measuring Questions: Relevance and its Relation to Entropy", *Bayesian Inference and Maximum Entropy Methods in Science and Engineering*, ed. C. J. Williams, AIP Conf. Proc. **659**, 517 (2004).

[33] J. E. Shore and R. W. Johnson, *IEEE Trans. Inf. Theory*, **IT-26**, 26 (1980); *IEEE Trans. Inf. Theory* **IT-27**, 26 (1981).

[34] J. Skilling, "The Axioms of Maximum Entropy", *Maximum-Entropy and Bayesian Methods in Science and Engineering*, eds. G. J. Erickson and C. R. Smith, (Kluwer, Dordrecht, 1988).

[35] A. Caticha, "Relative Entropy and Inductive Inference", *Bayesian Inference and Maximum Entropy Methods in Science and Engineering*, eds. G. J. Erickson and Y. Zhai, AIP Conf. Proc. **707**, 75 (2004) (arXiv.org/abs/physics/0311093).

[36] A. Caticha and A. Giffin, "Updating Probabilities", *Bayesian Inference and Maximum Entropy Methods in Science and Engineering*, ed. A. Mohammad-Djafari, AIP Conf. Proc. **872**, 31 (2006) (http://arxiv.org/abs/physics/0608185).

[37] S. N. Karbelkar, "On the axiomatic approach to the maximum entropy principle of inference," Pramana– J. Phys. **26**, 301 (1986).

[38] A. Renyi, "On measures of entropy and information," *Proc. 4th Berkeley Symposium on Mathematical Statistics and Probability*, Vol 1, p. 547-461 (U. of California Press, 1961).

[39] C. Tsallis, *J. Stat. Phys.* **52**, 479 (1988); "Nonextensive statistical mechanics: a brief review of its present status," (arXiv.org/abs/cond-mat/0205571). Critiques of Tsallis' non-extensive entropy are given in [40]; derivations of Tsallis' distributions from standard principles of statistical mechanics are given in [41].

[40] B. La Cour and W. C. Schieve, *Phys. Rev. E* **62**, 7494 (2000); M. Nauenberg, *Phys. Rev. E* **67**, 036114 (2003).

[41] A. R. Plastino and A. Plastino, *Phys. Lett.* **A193**, 140 (1994); G. Wilk and Z. Wlodarczyk, *Phys. Rev. Lett.* **84**, 2770 (2000).

[42] P. M. Williams, *Brit. J. Phil. Sci.* **31**, 131 (1980).

[43] P. Diaconis and S. L. Zabell, *J. Am. Stat. Assoc.* **77**, 822 (1982).

[44] A. Gelman, et al., *Bayesian Data Analysis, 2nd edition* (CRC Press, 2004).

[45] A. Giffin and A. Caticha, "Updating Probabilities with Data and Moments", *Bayesian Inference and Maximum Entropy Methods in Science and Engineering*, eds. K. Knuth, A. Caticha, J. L. Center, A. Giffin, C. C. Rodríguez, AIP Conf. Proc. **954**, 74 (2007) (http://arxiv.org/abs/0708.1593).

[46] A. Giffin, "Infering Diversity: Life after Shannon", *InterJournal of Complex Systems*, 2201 (2008) (http://arxiv.org/abs/0709.4079).

[47] E. Pielou. *Ecological Diversity* (Wiley, New York 1975).

[48] E. H. Simpson, "Measurement of Diversity", *Nature*, **163**, 688 (1949).

[49] A. Giffin, "Updating Probabilities with Data and Moments: A Complex Agent Based Example", *InterJournal of Complex Systems*, 2273 (2008) (http://arxiv.org/abs/0712.4290).

[50] A. Giffin, "From Physics to Economics: An Econometric Example Using Maximum Relative Entropy", *Physica A*, (2009) doi:10.1016/j.physa.2008.12.066, (http://arxiv.org/abs/0901.0401)

[51] E. T. Jaynes, *Probability Theory: The Logic of Science* (Cambridge, 2003).